\documentclass[12pt,preprint]{aastex}

\begin{document}
\title{FOURIER MODELING OF THE RADIO TORUS SURROUNDING SUPERNOVA 1987A}

\author{C.-Y. Ng\altaffilmark{1}, B. M. Gaensler\altaffilmark{1,a},
L. Staveley-Smith\altaffilmark{2,b}, R. N. Manchester\altaffilmark{3,a},
M. J. Kesteven\altaffilmark{3}, L. Ball\altaffilmark{3} and A. K. Tzioumis\altaffilmark{3}}
\altaffiltext{1}{Institute of Astronomy, School of Physics, The University of Sydney, Sydney NSW, Australia}
\altaffiltext{2}{School of Physics, The University of Western Australia, Crawley WA, Australia}
\altaffiltext{3}{Australia Telescope National Facility, CSIRO, Marsfield NSW, Australia}
\altaffiltext{a}{ARC Federation Fellow}
\altaffiltext{b}{Premier's Fellow in Radio Astronomy}
\email{ncy@physics.usyd.edu.au}

\begin{abstract}
We present detailed Fourier modeling of the radio remnant of Supernova 1987A,
using high-resolution 9\,GHz and 18\,GHz data taken with the Australia Telescope
Compact Array over the period 1992 to 2008. We develop a parameterized
three-dimensional torus model for the expanding radio shell, in which the
emission is confined to an inclined equatorial belt; our model also incorporates
both a correction for light travel-time effects and an overall east-west gradient
in the radio emissivity. By deriving an analytic expression for the two-dimensional
Fourier transform of the projected three-dimensional brightness distribution, we
can fit our spatial model directly to the interferometric visibility data. This
provides robust estimates to the radio morphology at each epoch. The best-fit
results suggest a constant remnant expansion at $4000\pm400$\,km\,s$^{-1}$ over the
16-year period covered by the observations. The model fits also indicate substantial
mid-latitude emission, extending to $\pm40^\circ$ on either side of the equatorial
plane. This likely corresponds to the extra-planar structure seen in H$\alpha$ and
Ly$\alpha$ emission from the supernova reverse shock, and broadly supports
hydrodynamic models in which the complex circumstellar environment was produced by
a progression of interacting winds from the progenitor. Our model quantifies the
clear asymmetry seen in the radio images: we find that the eastern half of the
radio remnant is consistently $\sim40\%$ brighter than the western half at all
epochs, which may result from an asymmetry in the ejecta distribution between
these two hemispheres.
\end{abstract}


\section{Introduction}
\object{Supernova (SN) 1987A} in the Large Magellanic Cloud (at a distance of
51\,kpc) was the nearest supernova observed in almost 400 years, and has enabled
detailed studies of supernova physics. The detection of neutrino bursts
indicated that the event was a core collapse, however SN 1987A was not a typical
example of a type II supernova in many aspects. Optical observations revealed a
triple-ring nebula surrounding the explosion \citep{bur95}. This indicates the inner
surface of the densest circumstellar material (CSM) that was ionized by the supernova
UV flash. Observations of the light echos provided a more comprehensive picture of the
structure \citep{sug05a,sug05b}. The CSM was found to have a peanut-like bipolar geometry,
inclined at 43\arcdeg\ to the line of sight.

The complex circumstellar environment can be understood as analogous to the interacting
stellar wind model, as originally proposed to explain planetary nebulae
\citep{kwo82}. The progenitor star of SN 1987A, \object{Sk $-69\arcdeg202$}, is believed
to have evolved from a red supergiant (RSG) to a blue supergiant (BSG) about 20000 years
before core collapse \citep{cro91}. After the transition, the fast, low density BSG wind
overran the slow, dense RSG wind and swept it into a circumstellar shell. Some models
suggest that the RSG wind could have been highly aspherical, due to a binary merger
\citep{mp07} or to equatorial wind compression \citep{col99}. Hydrodynamic simulations
of the subsequent interaction between the stellar outflow can reproduce the overall bipolar
CSM structure and can possibly explain the origin of the triple-ring nebula
\citep{bl93, ma95, mp07}.

At radio frequencies, an initial outburst from SN 1987A peaked on day 4 and then
decayed rapidly \citep{tur87}. The radio emission was redetected again in mid-1990
by the Molonglo Observatory Synthesis Telescope (MOST) \citep{tur90} and then by the
Australia Telescope Compact Array (ATCA) \citep{sta92}. Long term monitoring has
shown a monotonic increase in the radio flux since the reappearance \citep{sta07}.
This is believed to be synchrotron radiation from the continuous injection of nonthermal
particles as the blast wave expands into the CSM. In this general picture, electrons
are accelerated to ultrarelativistic velocities by the shock and then interact with the
magnetic field that is amplified by the Rayleigh-Taylor instability near the contact
discontinuity \citep{jn96}. Therefore the radio emission is expected to be generated
between the forward and reverse shocks. A detailed time-dependent model of diffusive shock
acceleration for SN 1987A was first discussed by \citet{bk92}. \citet{dbk95}
further improved the model by considering the accelerated ions that heat the upstream
plasma and modify the shock front. \citet{cd95} slightly modified the physical picture
and attributed the radio emission to the interaction between the ejecta and an H{\sc ii}
region located inside the optical ring. In this model, the H{\sc ii} region is a shell
of swept-up RSG wind photoionized by the BSG. The ongoing encounter between the shock
and CSM also predicts a steady increase in X-ray emission, which has been confirmed by
long term X-ray monitoring \citep{par07}. Moreover, it was suggested by \citet{cd95}
that the H{\sc ii} region would delay the encounter of the blast wave with the optical
ring to $2005\pm3$ CE. By then, there was predicted to be a drastic increase in the radio
flux.

From an energetic point of view, generation of the radio emission requires much less energy
compared to other wavebands such as optical and X-ray. Hence, radio data are a sensitive 
tracer of the shock interaction. As the supernova ejecta expand into the CSM, the radio
morphology will follow the evolution of the forward and reverse shocks and traces out the
global structure of the dense CSM. High resolution imaging observations with the ATCA have
been performed since mid-1991 \citep{sta93a}. Fig.~\ref{natural} shows the
diffraction-limited images at 9\,GHz. With an angular resolution of $0\farcs9$, these images
already reveal the spatially extended structure of SN 1987A. \citet{bri94} and
\citet{sta93a} tested and employed the super-resolution to further improve the resolution to
$0\farcs4$ and found that the remnant has a limb-brightened shell morphology
(Fig.~\ref{super}). Applying the same technique to observations in 1992-1995, \citet{gae97}
discovered two bright lobes in the shell. The lobes were aligned along the major axis of
the optical ring and the eastern lobe was about 1.2-1.8 times brighter then the western
lobe. The ATCA 18\,GHz upgrade provided twice the angular resolution as at 9\,GHz, and
confirmed all these structures in diffraction-limited images \citep{man05}. The 18\,GHz
image is well described by an inclined thick equatorial ring with the same orientation as
the bipolar CSM. Recently, Tingay et al.\ (2008, in preparation) performed a new VLBI
observation of SN 1987A. The 1.4\,GHz image similarly shows two lobes coincident with the
ones in the 9 and 18\,GHz images. In the hard X-ray band, a thick ring morphology has also
been observed \citep{par02}, suggesting a similar emission process as at radio frequency.
However, the optical hot spots, which are the inward protruding dense gas shocked by the
blast wave \citep{pun02,sug02}, show no correlation to the radio lobes in position
\citep{man02}.  Although the emergence of the optical hot spots coincided with the arrival
of the radio-producing shock at those regions \citep{bal01}, \citet{man02} indicate that
this had little effect on the morphology of the radio remnant.

As in many astrophysical observations, quantitative measurements of the physical
parameters require careful data modeling \citep[e.g.][]{nr08}. In our case, \citet{sta93b}
demonstrated that a robust way to obtain the remnant size is to model the visibility data
directly in the $u$-$v$ plane. They performed two dimensional (2D) Fourier transform of a
range of models including a spherical shell and a thin ring and then fitted them to
the visibility data to measure the remnant's expansion. \citet{gae97} employed a thin
spherical shell model in the fit and found a mean expansion velocity of $\sim3000\;
\mathrm{km\;s^{-1}}$ between 1992-1995. Later, \citet{man02} improved the modeling by
adding two point sources to the thin shell model to account for the eastern and western
lobes as observed. Recently, \citet{gae07} applied the thin shell model to fit all 9\,GHz
observations from 1992 to 2006 and suggested a gradual acceleration in the expansion
followed by a recent deceleration.

The ATCA imaging campaign has the longest time span for monitoring of SN 1987A at
any wavelength. With more than 17 years of observations from a single instrument,
we are now ready to improve the modeling. In this paper, we develop a better model
to account for more features in the remnant. We will model the observed east-west
asymmetry as well as the expected light travel time effects. The fits to all the
9\,GHz and 18\,GHz observations from 1992-2008 can then provide a better understanding
of the structure and evolution of SN~1987A. We describe the ATCA observations and
data reduction in \S\ref{s2}, then the detailed modeling and fitting procedures in
\S\ref{s3}. The results and discussion are presented in \S\S\ref{s4} and \ref{s5}.
Detailed mathematics of the model is listed Appendix~\ref{app}.

\section{Observations}
\label{s2}
The ATCA is a radio interferometer consisting of six 22-m dishes on a 6-km track, located
near Narrabri, NSW, Australia \citep{fra92}. The ATCA observations of SN~1987A discussed
here were performed at 9 and 18\,GHz with 6\,km array configurations. At 9\,GHz, the
source was observed approximately every six months with long integration times (6-12 hr)
at two simultaneous frequencies in the range 8.6 to 9.1\,GHz. In the observations, the
phase center was placed 10\arcsec\ south of the remnant to avoid any possible artifacts
\citep{eke99}. Detailed observational parameters of all the datasets used in this study
are listed in Table~\ref{tab1}. Two ATCA 18\,GHz observations are included in the analysis.
Each was taken one day before or after a 9\,GHz observation, thus allowing a direct
comparison between the two frequency bands. Except for the last epoch, a preliminary
analysis of all the 9 and 18\,GHz data has been presented in \citet{gae07}.

The data reduction was carried out with the \emph{MIRIAD} package \citep{stw95}. We
examined the data carefully for bad points and rejected scans during poor atmospheric
phase stability. From 1996 onwards, the source has sufficient signal-to-noise ratio
(S/N) to allow phase self-calibration, which was performed with a 5 minute solution
interval time. This time scale is long enough to integrate an appreciable S/N, but
is short enough to track the atmospheric phase variations. The phase self-calibration
results in a significant improvement in the fidelity of the complex gains for each
antenna. After the calibration, we shifted the phase center to the supernova site and
then averaged the visibility data in 5 minute intervals. Again, this is long enough to
obtain good S/N on each baseline, but is short enough to track the structure of the
source. We note that there is negligible loss in amplitude due to time-averaging
of the data, because the source is of small angular extent and has been shifted to the
phase center \citep{bs99}.

\section{Modeling} 
\label{s3}
\begin{figure}[!ht]
\includegraphics[angle=270, width=\textwidth]{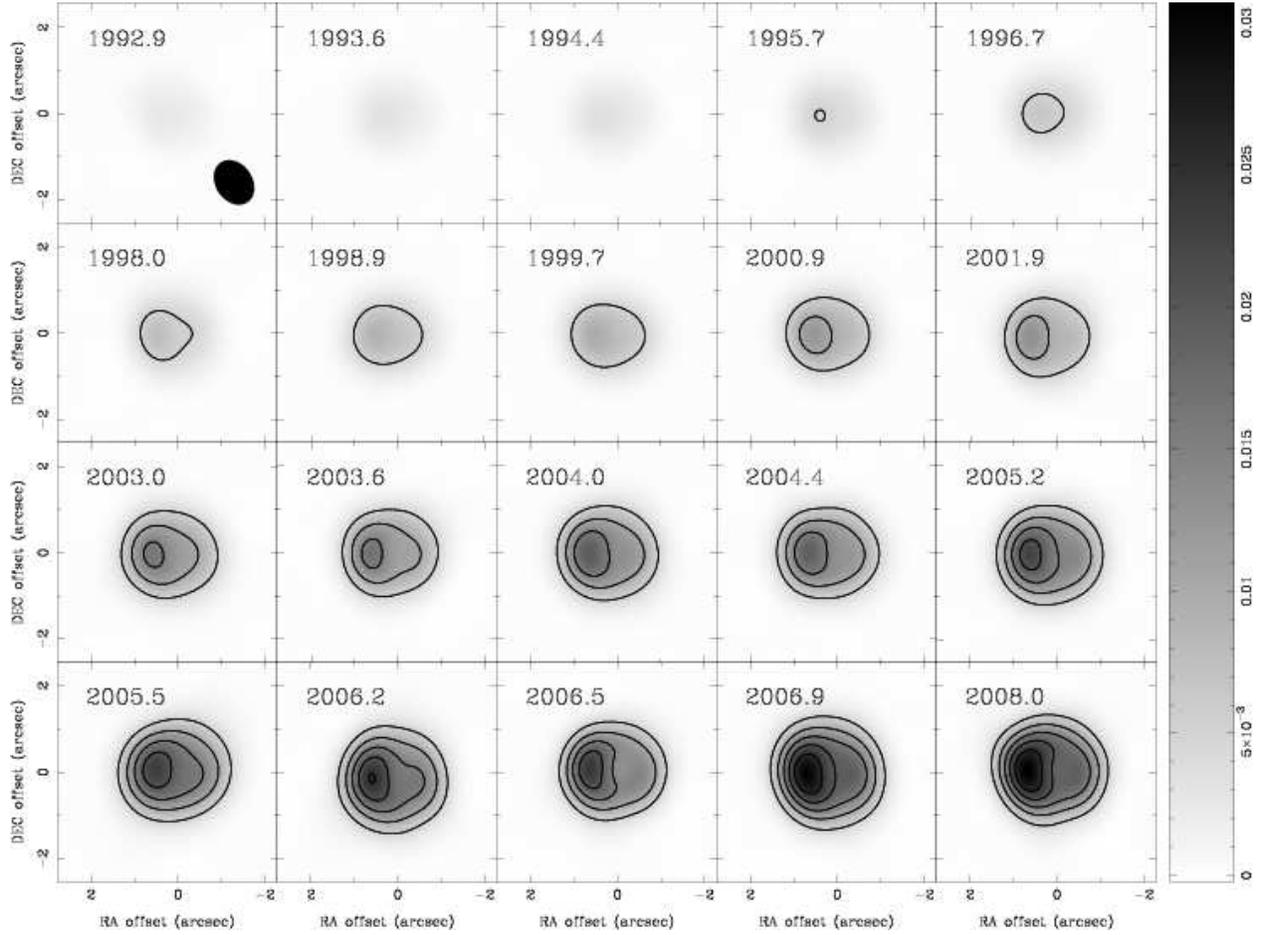}
\caption{Diffraction-limited 9\,GHz ATCA images of SN 1987A over the period 1992 to 2008.
 Some epochs have been averaged together to boost the S/N
(see Table~\ref{tab1}).
The scale is linear, ranging from --0.2 to 31\,mJy\,beam$^{-1}$. The contours are at
levels from 5 to 30\,mJy\,beam$^{-1}$ with 5\,mJy\,beam$^{-1}$ intervals. The synthesized
beam for the first epoch is shown in the first panel as an ellipse. The beams for the
other epochs vary slightly in orientation, but are similar in size.
\label{natural}}
\end{figure}

\begin{figure}[!th]
\includegraphics[angle=270, width=\textwidth]{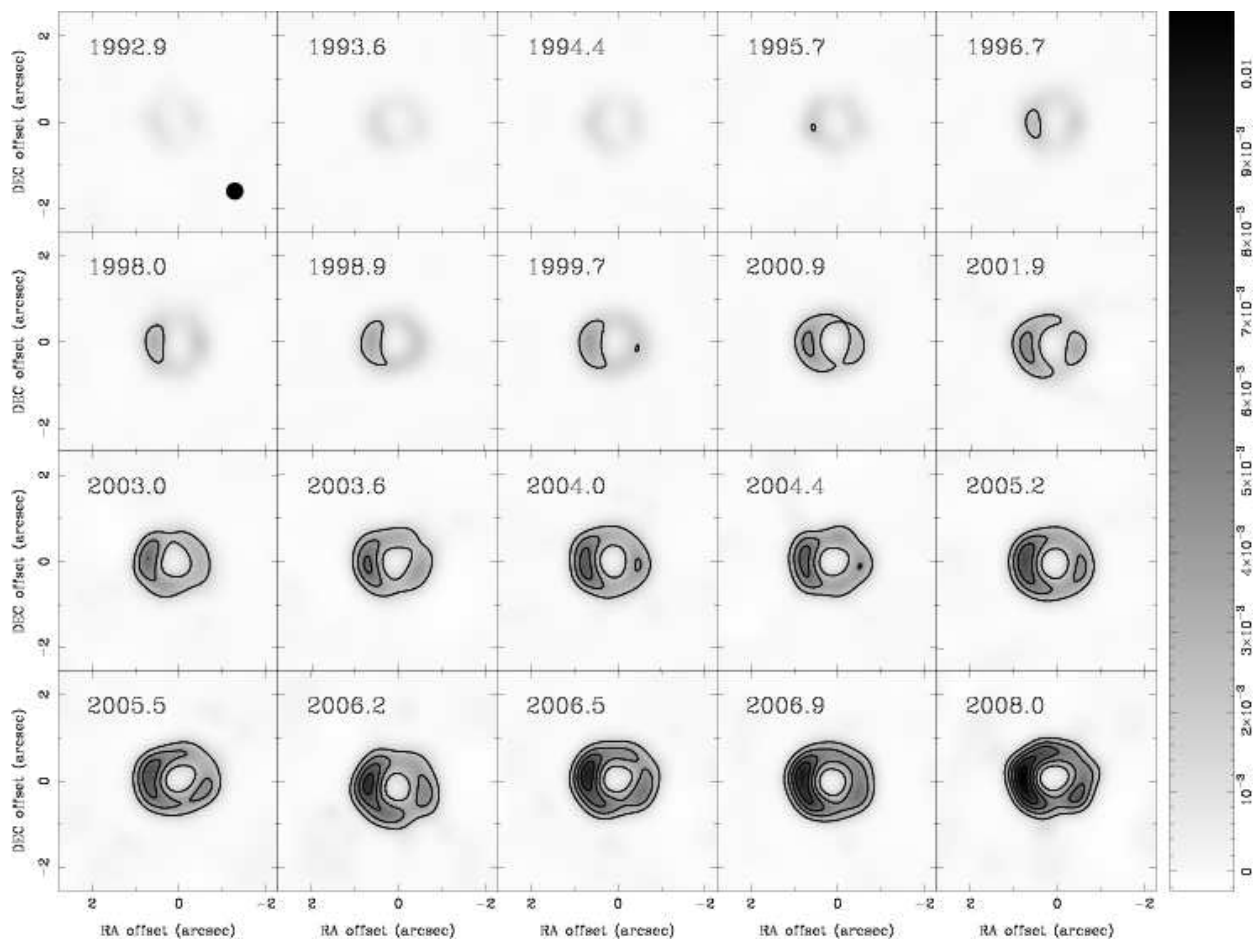}
\caption{Super-resolved 9\,GHz ATCA images of SN 1987A over the period 1992 to 2008.
 Similar to Fig.~\ref{natural}, some epochs have been
averaged to boost the S/N (see Table~\ref{tab1}). The scale is linear, ranging from
--0.2 to 10.8\,mJy\,beam$^{-1}$. The contours are at levels from 2 to 10\,mJy\,beam$^{-1}$
with 2\,mJy\,beam$^{-1}$ intervals. The synthesized beam, which is shown in the first
panel, is a circular Gaussian with FWHM 0\farcs4.\label{super}}
\end{figure}

Figures~\ref{natural} and \ref{super} show the diffraction-limited and super-resolved
images, respectively, of the 9\,GHz observations; some epochs were averaged together to
boost the S/N (see Table~\ref{tab1}). Following the procedures described by
\citet{gae97}, the dirty maps were first generated from the calibrated visibility data,
then deconvolved using a maximum entropy (MEM) algorithm \citep{gd78}. The MEM model was
then convolved with a diffraction-limited beam and a super-resolved circular beam with
FWHM $0\farcs4$ for Fig.~\ref{natural} and \ref{super}, respectively. Finally the models
were combined with the unscaled residuals and then regridded at a pixel scale of 0\farcs01.
The structure observed in the super-resolved 9\,GHz images is confirmed by the 18\,GHz
diffraction-limited images \citep[see Fig.~1 in][and Fig.~\ref{12mm} of this paper]{man05},
showing a morphology that is well-described by an inclined thick ring. Based on the surface
brightness at the center of the ring, \citet{man05} argued that the radio morphology sits
somewhere between a spherical shell and an equatorial ring.

Motivated by all these results, we below present a generic inclined torus model to capture
the overall structure of the radio emission. We emphasize that our model is a simplified
picture of the physical situation and that we have not attempted to reproduce every single
detail of the emission.  Instead, this study aims to obtain characteristic scales of the
emission, such as the size and height of the emitting region, and the evolution of these
parameters with time.

\subsection{Equatorial Belt Torus Model}
\label{s31}
The extra-planar structure of SN 1987A mentioned above obviously does not fit into the simple
picture of a spherical shell or a 2D ring, but is better described by a 3D torus with a
finite thickness. To incorporate these features, we propose an equatorial belt torus model,
i.e.\ a spherical shell truncated at arbitrary latitude. This gives a continuous variation
between a ring and a spherical shell geometry. Fig.~\ref{draw} illustrates this model. It is a
truncated optically thin shell with uniform emissivity and inclined to the observer line of
sight. The geometry is specified by the radius $R$ (averaged between the inner and outer
shells), the shock thickness $\delta$ (as a fraction of $R$) and the half opening angle
$\theta$. Additional parameters include the flux density $f_0$, center position $(x_0,y_0)$,
inclination $\zeta$ and position angle (P.A.) $\Psi$ on the plane of the sky. In the limiting
case of $\theta=0\arcdeg$, the torus is reduced to an equatorial ring, while $\theta=90\arcdeg$
gives a spherical shell. The east-west asymmetry of the remnant is modeled by a linear gradient
in the torus emissivity across the equatorial plane (not in the plane of the sky). Hence two more
parameters are introduced: the magnitude $g$ and the P.A. $\phi$ of the gradient. The former
is normalized to the fractional change in flux density at $R$, and the gradient direction is
defined from the brightest point towards the faintest point, with the P.A.\ measured from the
projected observer line of sight (see Fig.~\ref{draw}). Finally, we correct for light travel time
effects. Since the remnant flux is increasing, the north rim, which is about a light year closer
to the Earth than the south, appears brighter. Also due to the finite speed of light, expansion
of the remnant distorts the torus and shifts its center to the north. To derive corrections for
both effects, we employed linear approximations to the expansion and light curve locally at each
epoch, with the rate of change determined from the results of \citet{gae07} and \citet{sta07}.
We found that the north rim is at most 20\% brighter than the south as a result of the light
travel effects. On the other hand, the effects of expansion are only of second order, at the
level of less than 1\%.  Therefore, the latter will be ignored throughout the following
analysis. A detailed mathematical description of the torus model and light travel travel
time effects is given in Appendix~\ref{app}.

\subsection{Fitting Procedure}
\label{s32}
\citet{sta93b} showed that a robust way to estimate the remnant's geometry is to
model the visibility data directly in the $u$-$v$ domain, for which the measurement
errors are uncorrelated and the fitting is not complicated by the deconvolution
process. In order to do that, we first have to derive a 2D Fourier transform for
the inclined torus model. This was achieved by building up the model torus from
thin rings, which can be transformed to Fourier space analytically via the Hankel
transform \citep{bra00}. The linear gradients in the model, including the
asymmetry and light travel time correction, are transformed using the convolution
theorem and are turned into derivatives in the $u$-$v$ domain. Putting these
together, we obtain an analytic expression of the torus model under a Fourier
transform (see Appendix~\ref{app} for the detailed formulae).

Once the model is established, the fitting procedure is by the minimization of
\[ \chi^2 = \sum \left | \frac{\mathrm{vis}(u,v)-\mathrm{model}(u,v)} {\sigma(u,v)}
\right | ^2 \ ,\] where $\sigma$ is the statistical uncertainty associated with each
measurement. Given that the rms noise levels in our final images is close to the
thermal noise of the antenna receivers, we argue that other sources of errors in the
measurement, including the systematic errors due to the calibration or instrumental
effects are negligible. The $\chi^2$ minimization is carried out by the task
{\tt uvfit} in \emph{MIRIAD}, with modifications to the code for the torus model.
As discussed, the symmetry axis of the system has been well measured from optical
observations of the triple-ring system and of light echoes \citep{sug05b,pun07}.
Therefore, we fixed the orientation of the torus during the fit, by adopting the
values $\Psi=-7\fdg6$ and $\zeta=43\fdg4$ from \citet{pun07}. The uncertainties in
the fits were estimated through a bootstrapping method \citep{et93}, since the errors
in the fits are likely to be non-Gaussian. Following the procedure described in
\citet{pre92}, boostrap samples were generated by randomly drawing the visibility
measurements from an observation, with replacement. After the same number of
visibilities as the data have been drawn, the bootstrap replica is fitted with the
same procedure to obtain the torus parameters. For each observation, we fitted 2000
bootstrap iterations, and the 68\% intervals for each parameter's distribution are
quoted throughout as the $1\sigma$ uncertainty. This gives the projection of
multi-dimensional errors that accounts for the correlations between fitting parameters.
As the S/N in the observations increases towards the later epochs, the parameter
uncertainties due to measurement errors become negligible and the systematic errors
in the modeling dominate. In the real world, the detailed structure of SN 1987A is much
more complicated than what our simple model can capture, resulting in systematic errors
in the fits. We have attempted to quantify this term for the radius, which is the
parameter we are most interested in. Since the exact structure of the remnant is
unknown, the radius measurement is always model-dependent. Similar to \citet{man02}, we
fitted the 9\,GHz observations to various plausible models, including a spherical shell
and a shell plus two point sources. In general, these fits obtain slightly different
values of $R$ and hence the variation provides a handle of the systematic errors in
the modeling.

\section{Results}
\label{s4}
\subsection{9\,GHz Observations}
\label{s41}
The best-fit torus parameters are listed in Table~\ref{tab2} and plotted in Fig.~\ref{plot}.
Most of the fits to the ATCA 9\,GHz data are robust, except for a few epochs with very
large $\chi^2$ values that are possibly due to poor calibration. The fits are generally
better towards the late epochs, however as shown in Table~\ref{tab2}, the typical reduced
$\chi^2$ value is $\sim1.5$. This is still relatively large, suggesting that our model is
not a perfect fit to the observations, and that there are unmodeled features in the
data which contribute to the systematic residuals. This result is not unexpected since the
model aims to capture only the characteristic features of the emission, rather than
reproducing every small-scale structure. Nevertheless, our model does provide a better
description to the data than a simple thin spherical shell model used in previous studies
\citep{gae97,gae07,man02}. Fig.~\ref{vis} plots the visibilities of the data and models along
two orthogonal axes in the $u$-$v$ domain. The difference between a torus and a spherical
shell models is obvious, in particular along the $u$-axis, because this direction
corresponds to east-west in the sky, where the radio emission shows large asymmetry. The plots
clearly indicate that the asymmetric torus model is a better fit over a simple shell, and, at
least to the first order, well describes the east-west asymmetry. On the other hand, we note
that a similar plot with azimuthally averaged $u$-$v$ data \citep[Fig.~6 in][]{gae07} masks the
discrepancy between the data and a simple shell model. To provide a quantitative comparison
between different models, we have reproduced the fits with a thin shell model and a thin
spherical shell plus two points model as in \citet{man02}. Table~\ref{tab3} shows that
the torus fit always provides lower $\chi^2/\nu$ values than the shell fit. The shell plus
two points model and the torus model have comparable goodness-of-fit, but the former
requires two more fitting parameters. Therefore, we conclude that our torus model is a
substantial improvement over the simple models used previously.

Although we did not attempt to provide a physical model to interpret the observed
radio light curve, we note that the fluxes obtained from the fits are consistent with the
values reported by \citet{sta07}. In the following discussions, we will concentrate on the
torus geometry.

Fig.~\ref{model} shows the best-fit model for epoch 2008.0, with the left panel
illustrating the model at very high spatial resolution. The images were generated in the
$u$-$v$ domain with complete uv coverage, then Fourier transformed back to the image
plane. The right panel shows the image convolved with a 0\farcs4 FWHM Gaussian, the same
as the super-resolved beam used in Fig.~\ref{super}. The inclined torus model has
successfully captured the aspect ratio of the overall structure, and has also reproduced
the two lobe features, thus confirming that they are due to projection of the optically
thin equatorial belt.

A better way to compare between the model and data is to generate an image of the model
with identical $u$-$v$ sampling as the corresponding observation. Then the sampling
effects will appear identically in both images, allowing a direct comparison.
Fig.~\ref{fits} \& \ref{2008} show the images of the data compared to the best-fit models for
selected epochs. The models were generated with the approach described above and then imaged
using the same procedures as mentioned in \S\ref{s2}. However, since the MEM algorithm is
not applicable to images with negative values, we did not attempt to deconvolve the
residual maps. We emphasize that the residual maps presented
here only illustrate a qualitative comparison between the data and the model, rather than a
detailed quantitative description of the underlying difference. The most obvious feature in
the residuals is the flux deficit of the model torus in the southeast and northwest, and
the excess emission at the center. The discrepancies are due to limitations of our
relatively simple model. As indicated by the super-resolved images, the eastern lobe in the
data is not reflection symmetric about its brightest point, but extends further to the
south. This cannot be easily described by the model, which approximates the asymmetry by
a linear gradient. Although less obvious, a similar case occurs for the western lobe as
well. On the other hand, the excess flux at the center of the model torus is due to the
abrupt cutoff of the high latitude emission in the model. Since the inclination angle
$\zeta$ is fixed during the fit, the ellipticity depends solely on the opening angle
$\theta$. A large $\theta$ is required by the fit in order to match the overall geometry,
hence resulting in excess emission filling the interior. In real life, the radio emissivity
is likely to be latitude dependent and to decay smoothly at high latitude, providing a
slightly different overall ellipticity. A more realistic model should incorporate all these
effects to address the above discrepancies.

\subsection{18\,GHz Observations}
The best-fit model for the 18\,GHz observation of 2003 July is shown in Fig.~\ref{12mm},
with the data and model visibilities plotted in Fig.~\ref{vis12}. The diffracted-limited
images were deconvolved the CLEAN algorithm. Similar to Fig.~\ref{fits} \& \ref{2008},
no deconvolution has been applied to the residual map in order to avoid running the algorithm
into noise. As shown in Table~\ref{tab2}, the best fits have small $\chi^2/\nu\approx1$, which
are due to large measurement uncertainties (see Fig.~\ref{vis12}). The best-fit parameters
are plotted by the open circles in Fig.~\ref{plot}. They are all consistent with the 9\,GHz
results, except that $\theta$ is slightly smaller at 18\,GHz. While the discrepancy in $\theta$
seems to suggest a frequency-dependent emission height that implies different physical conditions
between the equatorial plane and at higher latitude, it could also be the result of systematic
errors in the fits which we do not fully understand. As the remnant is getting brighter,
future observations will help to investigate this problem. Following \citet{man05}, slices
through the 18\,GHz diffraction-limited images along different P.A.\ were plotted in
Fig.~\ref{slice}. In general, the two peaks in the model match well to the observation. The
only exception is at 150\arcdeg, where the data are $\sim25\%$ brighter than the model in the
east. As discussed in \S\ref{s41} above, this is attributed to the unmodeled southern extension
of the lobe. The plot also illustrates the excess emission near the center of the model torus,
again consistent with the 9\,GHz results.  

\section{Discussion}
\label{s5}
\subsection{Remnant Size}
In general, the size of the torus is well-determined from the fits and is consistent with
that determined in previous studies \citep{gae97,man02,gae07}. The best-fit radii from the
torus and shell fits are comparable, but those from the shell plus two points model are
always slightly larger. A similar discrepancy was also reported by \citet{man02}, suggesting
that the radius measurement is somewhat model-dependent. The average discrepancy is $\sim5\%$
for the later epochs, which is attributed to the systematic errors in the modeling as mentioned
in \S\ref{s32}. In the following discussion, we assigned a 5\% systematic errors for $R$. Hence
the quadrature sum of these and the statistical errors obtained from bootstrapping provides
the total uncertainties in $R$, which are shown by the dotted error flags in Fig.~\ref{plot}.

Compared to the measurements published for the X-ray shell \citep{par07}, the radio torus
is larger by 10-15\% which is too large to be accounted for by the measurement uncertainties.
However, \citet{gae07} analysed the Chandra data in the Fourier domain and found that the
X-ray and radio shells agree to better than 1\% in size. Hence they argued that the
discrepancy is due to different measuring techniques, rather than any physical difference.
At optical wavelengths, \citet{mic98} reported that the reverse shock was located at 70\% of
the inner circumstellar ring around day 3900. This gives a radius of 0\farcs64, which is
slightly smaller than the torus size at that epoch. From the steady increase in the radio
light curve, \citet{man05} argued that the radio-emitting electrons could be filling the
region between the shocks. Although the thickness $\delta$ is not too sensitive in the
fits and generally very small, this is still consistent with the prediction. With a density
profile $\rho(r,t) \propto t^{-3}(r/t)^{-9}$ in the supernova ejecta \citep{ek89},
numerical simulations suggest that during the early stage of the SNR evolution, the
forward and reverse shocks are separated only by 10-20\% in radius \citep[e.g.,][]{tm99}.
For \object{SN 1993J}, \citet{bar07} reported a shell thickness of 30\% of the inner shell
radius. Despite the above results assuming a spherical symmetry, if $\delta$ in our case is
of a similar order, given that the whole remnant is barely resolved in the 9\,GHz
observations, $\delta$ likely remains unresolved. This also supports the conclusion of
\citet{gae07} that both thick and thin shell models provide equally good fits to the data.

\subsection{Rate of Expansion}
At the earliest epoch, the remnant has a radius of about 0\farcs6. As \citet{gae97} pointed
out, this implies a minimum mean expansion rate of $30000\;\mathrm{km\;s^{-1}}$ during the
period 1987 to 1992, in the free expansion phase. This expansion was then followed by a
drastic deceleration to $\sim3000$\,km\,s$^{-1}$ at around the time the radio emission
re-emerged \citep{gae97}. To determine the expansion since day 1800, we employed a linear
fit to $R$ for all the 9\,GHz observations and obtained an expansion velocity of
$v=4000\pm400\;\mathrm{km \;s^{-1}}$ over the period 1992-2008. The small $\chi^2/\nu=0.2$
in the fit indicates that the measurement uncertainties are over-estimated.  On the other
hand, if only the statistical errors are considered, the result is similar
($v=4150\pm30\;\mathrm{km\;s^{-1}}$), but with smaller errors and a much larger
$\chi^2/\nu=9.0$. \citet{gae07} performed a quadratic fit to the remnant size and suggested
an acceleration in the expansion. We also found a similar trend in our results. However, a
$F$-test suggests that the quadratic fit is not statistically significant: there is a 20\%
probability that the smaller $\chi^2$ in the quadratic fit would occur solely by chance.
Therefore, we conclude that the simpler model (i.e.\ linear fit) is preferred and adopt the
linear expansion $v=4000\pm 400\;\mathrm{km\;s^{-1}}$ in the following discussion. This
value lies between $v=4700\pm100\;\mathrm{km\;s^{-1}}$ and $\sim3700\;\mathrm{km\;s^{-1}}$
obtained by \citet{gae07} and \citet{man02} respectively using a thin shell model, and
agrees with the prediction $v=4100\;\mathrm{km\;s^{-1}}$ from hydrodynamic simulations
\citep{bbm97}. It is also similar to the expansion rate of the reverse shock at
$v=3700\pm900\;\mathrm{km\;s^{-1}}$ reported by \citet{mic98}. On the other hand, the X-ray
expansion is apparently much faster before day 6200. Early measurements showed an X-ray
expansion of $v=4200\;\mathrm{km\;s^{-1}}$ \citep{par04}, but \citet{par07} recently
suggested acceleration to a higher velocity of $v\sim6000\;\mathrm{km\;s^{-1}}$ over the
period covering 5000-6000, before the expansion decelerates to $v=1400\;\mathrm{km\;s^{-1}}$
near day 6200. This could indicate that the X-ray emission is located differently as the
radio emission and much closer to the forward shock.

A comparison between the radio and optical observations suggests that the supernova blast
wave may have started to encounter the dense CSM in the past few years. The optical ring,
the emission from which represents fluorescence after ionization of the CSM by the supernova
UV flash, indicates the inner boundary of the dense swept-up CSM. As seen by the \emph{HST},
the central equatorial ring has a radius of 0\farcs81-0\farcs85 \citep{bur95, pla95}.
Accounting for the uncertainties in both the radio and optical measurements, our results
suggest that the radio emission could have reached the optical inner ring around day
6000-6500, hence the forward shock could have started to expand into the dense CSM as early
as 2004. This falls nicely within the predictions by \citet{cd95}, \citet{gae97} and
\citet{man02}, but disagrees with other estimates such as those of \citet{lms94} and
\citet{bbm97}. The X-ray observations also provide evidence for the encounter. The rapid
increase in the soft X-ray flux near day 6000-6200 \citep{par05} and the drastic
deceleration of the X-ray expansion around day 6200 \citep{par07} reveal that the forward
shock has encountered a steep gradient in the ambient density. We expect a similar
deceleration in the radio expansion to occur very soon.

\subsection{Emission Height} 
In addition to the radius, the opening angle $\theta$ is also an important parameter that
probes the CSM geometry. The best-fit values of $\theta$ shown in Fig.~\ref{plot} are
substantial, indicating a significant level of radio emission from mid-latitudes. This
extra-planar structure implies the spread of dense CSM over large regions on either side of
the equatorial plane, and suggests that the prominent optical ring is merely the densest
part of a more complicated 3D structure. In the later epochs, $\theta$ stays nearly
constant around 40\arcdeg. This gives an inferred emission height up to $\sim5\times
10^{17}\,$cm from the equatorial plane. Our result indicates a radio emission geometry very
similar to that of the reverse shock. \citet{mic98,mic03} reported a rapid decrease of
Ly$\alpha$ and H$\alpha$ emission beyond $40\arcdeg$ and hence deduced that the reverse
shock is mostly confined within $30\arcdeg$ of the equatorial plane. Our fitting results
also provide the characteristic structure of the CSM, which can be compared to hydrodynamic
simulations such as those of \citet{bl93}, \citet{ma95} and \citet{mp07}. In those simulations,
the exact shape of the contact discontinuity between the BSG and RSG winds is highly
model-dependent.  Some models suggest a rather constant opening angle over a large range of radius
\citep[e.g.\ Fig.~3 of][]{ma95}, which is similar to our case. Nevertheless, all the
simulations indicate that the overall geometry of the swept-up CSM is peanut-like, with a
dense equatorial ring at the waist. This picture predicts that the radio emission will
extend with time to higher latitude. Although the expected trend is not found in the
current data, we anticipate $\theta$ to increase significantly in future observations
according to the above models. Eventually the shock will reach the dense CSM in the polar
region and the radio morphology will be more circular. Our results do not show any polar
protrusions as suggested by \citet{blc96}. If the forward shock continues to expand at a
similar rate, it may take another 10-20 years to reach the polar region before we can
verify these predictions. As a note, Fig.~\ref{plot} suggests some hints that $\theta$ is
slightly larger in the early epochs. If this is real, it may indicate the encounter of
ejecta with dense regions at higher latitude at the early stages. Along with the shock
expansion, the radio emission is then gradually dominated by other regions closer to the
equatorial plane, resulting in a gradual decrease in $\theta$.  

\subsection{Asymmetry}
The super-resolved images in Fig.~\ref{super} clearly show that the eastern lobe is brighter
than its western counterpart. This feature has also been reported in the X-ray observations
\citep{bur00,par02,mic02}. In our torus model, the asymmetry is quantified in terms of a
linear gradient with direction $\phi$ and magnitude $g$. Fig.~\ref{plot} shows a trend of
increase in $\phi$ prior to 1998. Although this occurred at a similar time as the optical
hot spots first appeared, comparing $\phi$ to the hot spot positions shows that they do not
coincide. Therefore we arrive the same conclusion as \citet{man02} that the detailed
morphology of radio and optical emissions are not correlated. The only similarity is that
they both brightened first in the eastern half, indicating a global asymmetry along the
east-west direction. On the other hand, the variation in $\phi$ may be associated with the
slight decrease in $\theta$ around that period.  This could suggest that the blast wave has
encountered different dense regions before and after $\sim1998$. In the late epochs, $\phi$
stays around 110\arcdeg, which corresponds to a P.A.\ of 96\arcdeg\ on the plane of the sky.
This matches the bright X-ray lobes located at P.A.\ 90\arcdeg\ and 270\arcdeg\ as reported
by \citet{par02}.

Although there were indications from earlier data that the asymmetry increases with time
\citep{gae97,man02}, the longer data span presented here does not support this. From 1995
onward, $g$ stays at a constant level of $\sim40\%$. This gives a flux ratio of 1.5 between
the two halves of the torus at the equatorial plane. The projection onto the plane of the
sky reduces this ratio slightly. Thus as seen by an observer, the eastern lobe appears to
be $\sim1.4$ times brighter than the west. In order to confirm this result, we compared to
the degree of asymmetry inferred from the shell plus two points fits. The surface
brightness ratio between the eastern and western lobes is listed in Table~\ref{tab3}; again
this does not show any significant long-term variations, but suggests a rather constant
flux ratio of about 1.3, which is consistent with the torus estimate. Hence we conclude
that the surface brightness of the eastern half is consistently 30-40\% larger than the
west. This is very close to the 40\% asymmetry in the reverse shock \citep{mic03}, also
similar to the 15-50\% asymmetry in the 0.3-6\,keV X-ray band \citep{mic02}. While the
exact origin of the east-west asymmetry still remains unclear, it seems likely that the
asymmetry either lies in the initial velocity of the ejecta, which would imply an
asymmetric explosion, or in the density of the CSM \citep{gae97,sug02}. In principle the
evolution of the radio morphology can help distinguish these two scenarios. If the eastern
lobe expands faster, it will cause a systematic shift of the torus center towards the east.
Unfortunately, the phase self-calibration process removes the absolute astrometric
information in the data, precluding a direct comparison between epochs. Nonetheless, the
shell plus two points fit can provide some insights into the relative positions of the
features. Although this model is less preferable than the torus model, both of them obtain
a comparable goodness-of-fit, therefore we believe that the former can still provide a
sensible description of the location of the lobe centroids. As listed in Table~\ref{tab3},
the offsets of the point sources from the shell center increase at a similar rate as the shell
radius.  This would not be the case if the CSM were denser towards the east, since the higher
density would lead to a larger deceleration for the eastern lobe. In the other scenario, if
the ejecta in the east had a higher velocity, then it would arrive at the equatorial ring first
and would start to decelerate. Hence we expect the ratio of offsets to vary in future
observations.  However, we emphasize that this is only a very rough estimate. In particular,
if the shell plus two points model is not a good description of the physical structure, excess
flux from the lobes will shift the center of the underlying shell, introducing large systematic
errors in the positions.

\subsection{Limit on Radio Flux of a Central Source}
In spite of extensive searches in radio and optical wavelengths, no pulsar has yet
been found within the supernova remnant \citep{man07}. Our fitting scheme can also
provide a detection limit on any possible central object. We applied the same fitting
procedure using the torus model with a point source fixed at the center. The flux density
of the point source was held fixed during the fit. We adjusted its level until the change in
the best-fit $\chi^2$ value exceeded a certain limit. To begin with, we applied the fit to
the 1996 July observation, in which the remnant has high S/N but is still not too bright
to preclude a sensitive search. The fit gives a 3$\sigma$ upper limit of 0.3\,mJy. Applying
the fit on several other epochs obtained very similar results. Hence we conclude that this
is the 3$\sigma$ detection limit on any unresolved central source, including a neutron star
or a compact pulsar wind nebula, for the ATCA observations at 9\,GHz. As a comparison,
the most sensitive limit on pulsation at this frequency yields an upper flux limit of
0.058\,mJy \citep{man07}.

\subsection{Comparison to Other Radio Supernova Remnants}
Among about ten resolved radio supernovae (RSNe) \citep[see][]{bie05}, SN 1987A shows a
very different characteristics in the asymmetry, morphology and expansion as compared to
the others. For example, SN 1993J in M81, which is the best studied RSN besides SN 1987A,
was found to be circular within 3\% over 7 years of VLBI observations. Based on a statistical
argument, \citet{bbr03} suggested that the emission is intrinsically spherical. On the other
hand, only two other RSNe show large deviation from a circular shell. VLBI imaging of
\object{SN 1986J} in NGC 891 revealed a distorted shell structure with a hot spot
\citep{bie04}. \citet{per02} suggested this could be due to protrusions in the progenitor's
wind \citep{blc96}, or jet-induced core collapse \citep{kho99}. Another example of a
non-spherical RSN is \object{SN 2004et} in NGC 6946. A recent study by \citet{mar07} interpreted
the asymmetry as either a shell with two hot spots, which could be similar to the case of
SN 1987A, or an anisotropically expanding shell with a protrusion.

The main distinction between SN 1987A and other RSNe lies in its rate of expansion. As
discussed, the expansion of SN 1987A must have undergone a rapid deceleration in the first
1800 days.  In contrast, for all other RSNe with detected expansion, their sizes are
well-approximated by a power-law $R \propto t^m$, with the deceleration parameter
$m\sim0.7-1$ \citep{bie05}. The most extreme example is \object{SN 1979C} in M100, which
has shown nearly free expansion ($m\approx1$) for over 22 years \citep{bb03}. The unusual
morphology and evolution of SN 1987A can be largely attributed to its complex circumstellar
environment. Thus, this extraordinary event provides a unique example in the study of
supernova shell dynamics, stellar wind interactions and stellar evolution.

\section{Conclusions}
In this study, we have developed a scheme to model the interferometeric $u$-$v$ data
for any 3D object with axisymmetric structure. An application to ATCA 9 and 18\,GHz
observations of SN 1987A has improved the spatial modeling of the radio remnant. An
inclined equatorial belt torus model in 3D was employed, with the incorporation of
light travel time effects and of an overall east-west asymmetry. The projected
brightness distribution of the model is transformed into 2D Fourier space analytically,
then is fit directly to the visibility data. The torus model successfully captures
the characteristic features of the radio morphology and provides substantial improvements
over the simple spherical shell fit used in previous studies. The best-fit torus geometry
shows significant extra-planar emission extending to mid-latitudes $\sim40\arcdeg$, and a
40\% flux difference between the eastern and western lobes. Evolution of the shell radius
with time implies a constant expansion rate of $4000\pm400$\,km\,s$^{-1}$ over the period
1992-2008 covered by the observations; the radio remnant is probably now starting
to encounter the optical inner ring. Our results suggest a torus geometry and expansion
speed both very similar to those seen in H$\alpha$ and Ly$\alpha$ emission, indicating that
the radio synchrotron emission originates between the supernova forward and reverse shocks.
In summary, the torus fitting provides a powerful tool to probe the 3D structure of the
radio emission, showing that the overall picture of the dense CSM is consistent with the
hydrodynamic simulation of the progenitor star's interacting winds. Future evolution of
the radio morphology can help differentiate between specific evolutionary histories for
this system, and can address the apparent east-west asymmetry in the ejecta distribution.


{\it Facilities:} \facility{ATCA ()}

\acknowledgments 
We thank Shami Chatterjee for useful discussions.
The Australia Telescope is funded by the Commonwealth of Australia for operation
as a National Facility managed by CSIRO. B.M.G. and R.N.M. acknowledge the support
of the Australian Research Council.

\appendix
\section{Two Dimensional Fourier Transform of an Inclined Torus Model}
\label{app}
In this section, we give a mathematical description of an inclined torus model
and derive an analytic expression for its Fourier transform. Although a specific example
of a torus model is considered here, the expression could be easily generalized to any
objects with axisymmetry.

\begin{figure}[!ht]
 \begin{minipage}[c]{0.32\textwidth}
  \centering
  \includegraphics[width=\textwidth]{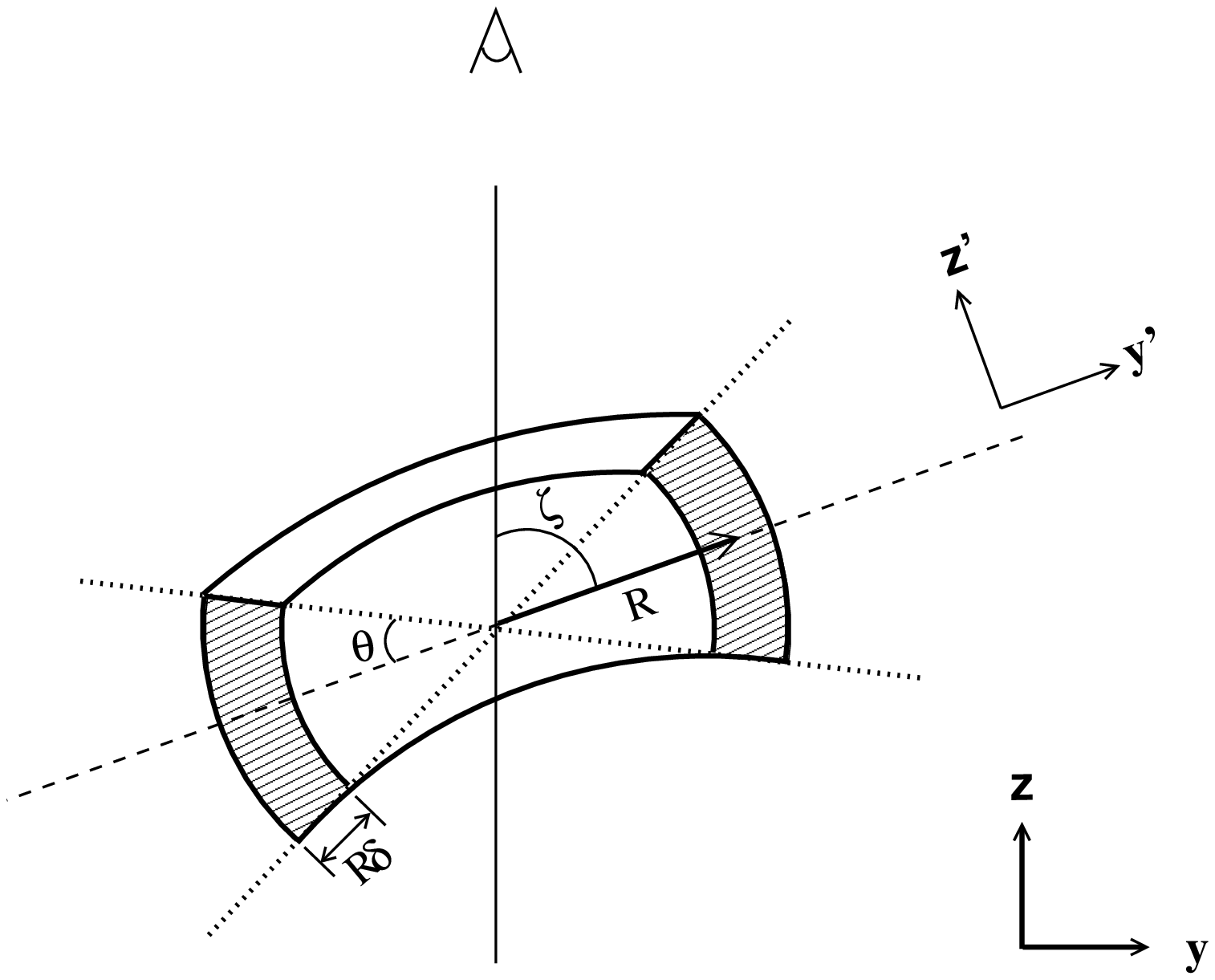}
 \end{minipage}
 \begin{minipage}[c]{0.32\textwidth}
  \centering
  \includegraphics[width=0.8\textwidth]{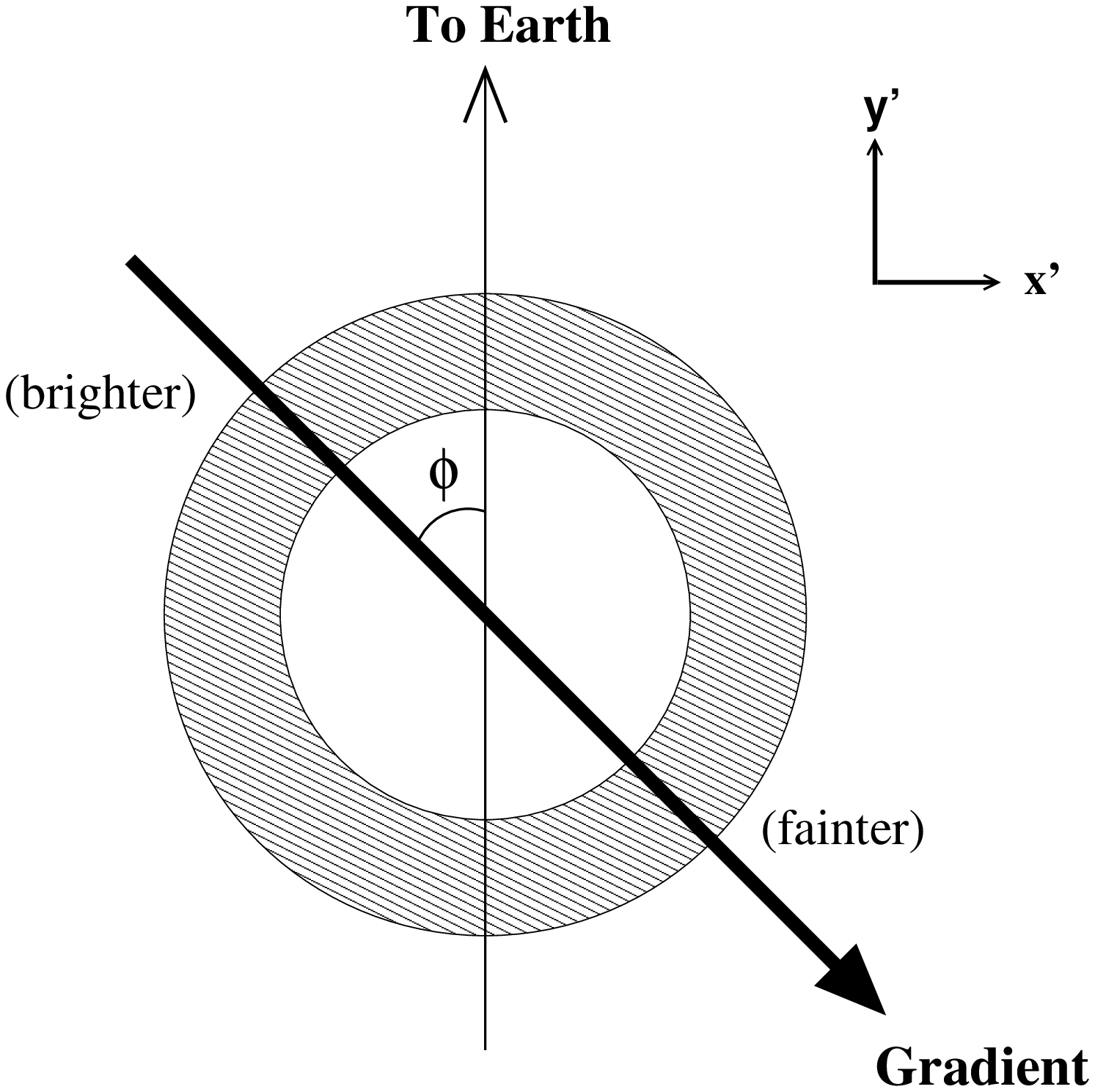}
 \end{minipage}
 \begin{minipage}[c]{0.32\textwidth}
  \centering
  \includegraphics[width=\textwidth]{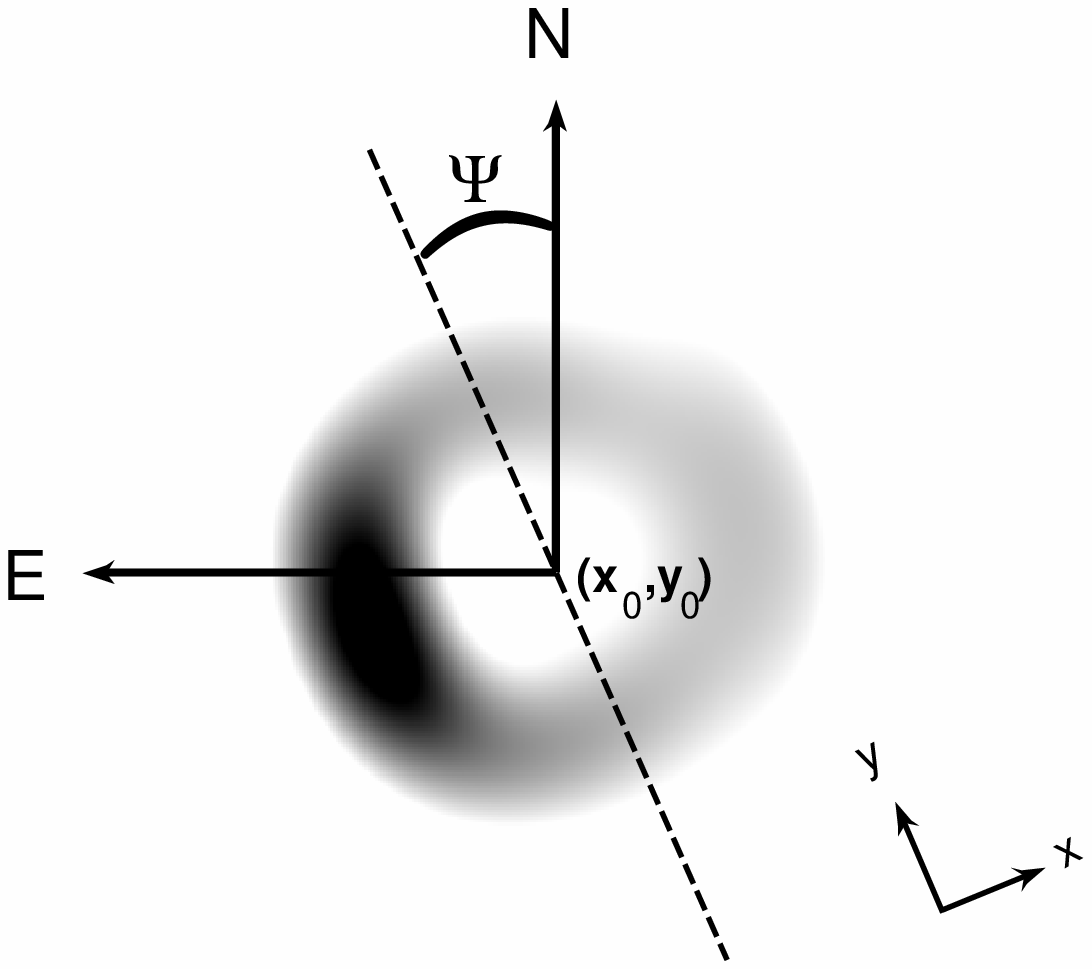}
 \end{minipage}

\caption{Schematic diagrams illustrating the geometric parameters
in the torus model. \emph{Left:} cross-section of the torus model, as
viewed perpendicular to the observer line of sight.
\emph{Middle:} the torus in the equatorial plane showing a linear gradient,
defined from the brighter side toward the fainter side.
\emph{Right:} A model torus, which is generated with parameters similar to
that of the best-fit model for the most recent epoch (2008.0), projected
onto the image plane (i.e.\ the plane of the sky).\label{draw}}
\end{figure}

\begin{figure}[!ht]
\plotone{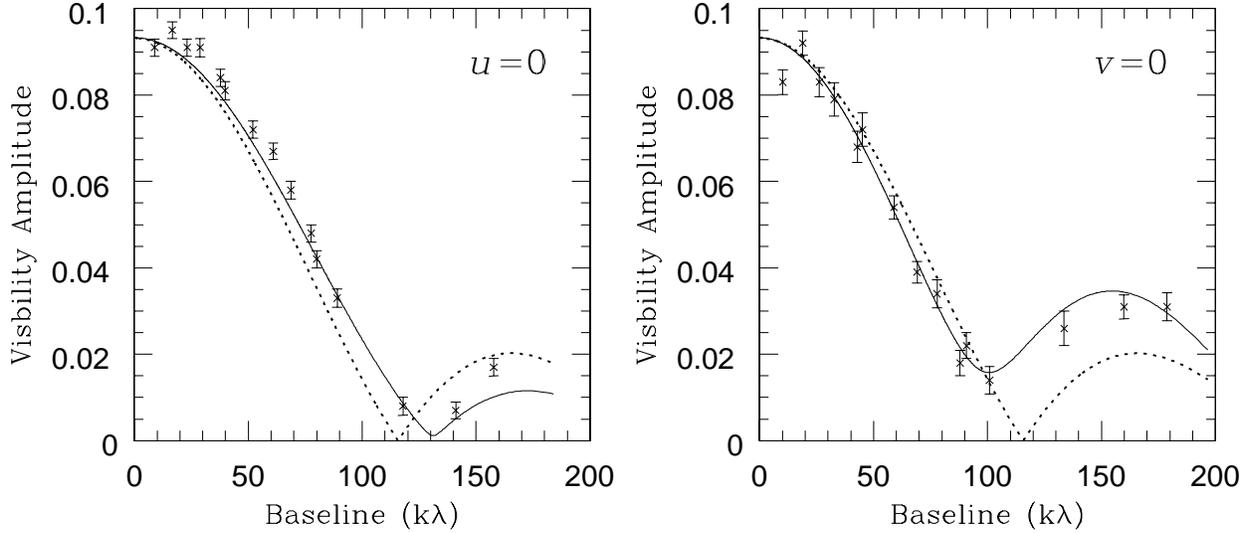}
\epsscale{1.0}
\caption{Amplitude of the visibility data in the most recent epoch (2008.0)
of ATCA 9\,GHz data compared to the best-fit torus and thin spherical shell models.
The torus and shell models are plotted by the solid and dotted lines respectively.
The two panels show slices through the $u$-$v$ domain along the axes $u=0$ and $v=0$.
\label{vis}}
\end{figure}

\begin{figure}[!ht]
\epsscale{0.8}
\plotone{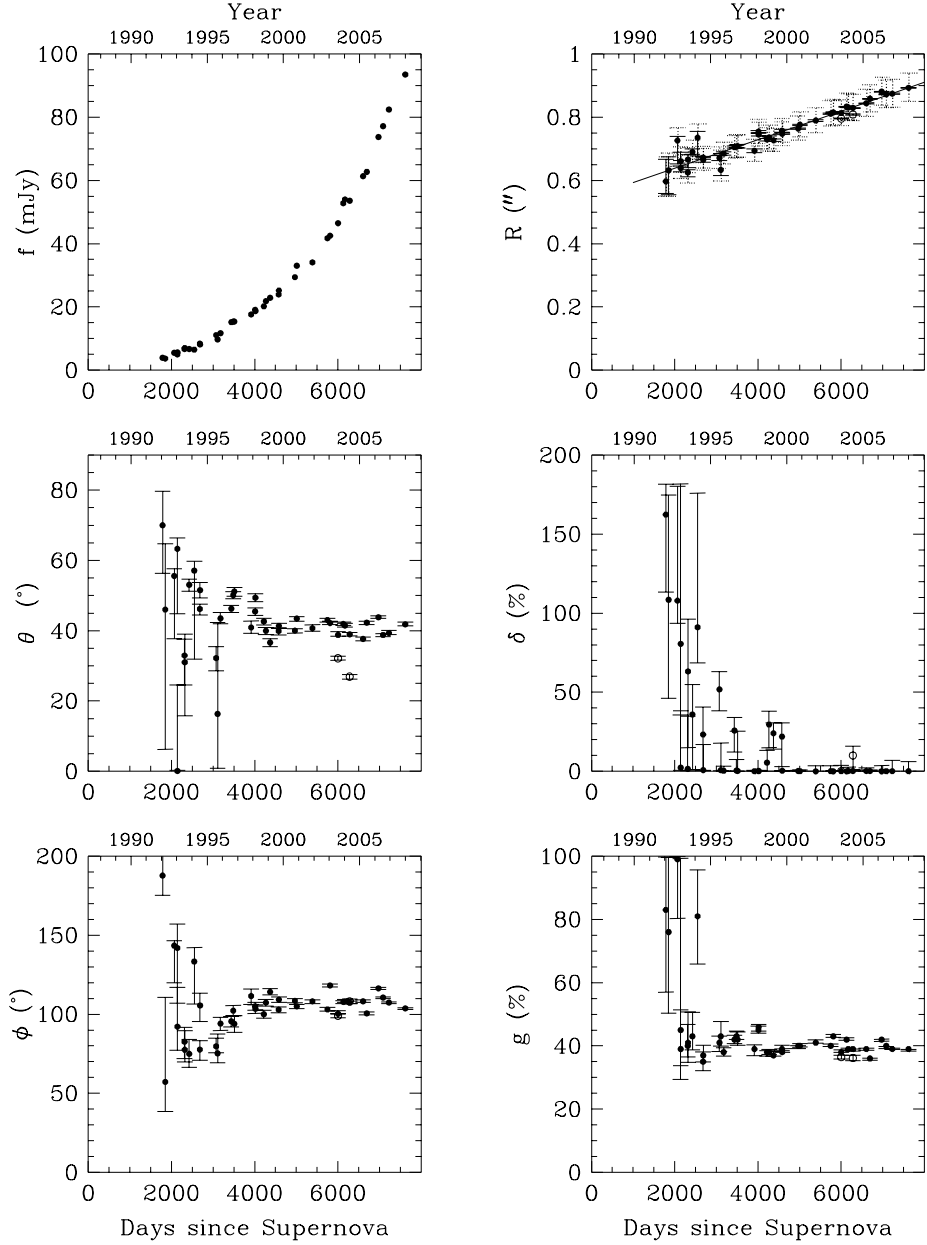}
\caption{The best-fit flux density ($f$), radius ($R$), opening angle ($\theta$),
shock thickness ($\delta$), gradient angle ($\phi$) and gradient magnitude ($g$) at
each epoch, obtained from the equatorial belt torus model as described in \S\ref{s31}.
The results for the 9 and 18\,GHz observations are plotted as filled and open circles
respectively. Note that we did not plot the flux density for the 18\,GHz observations,
since the absolute flux scale has not been fully calibrated. The uncertainties in flux
density is negligible. The dotted error flags in $R$ indicate the total uncertainties,
which are the quadrature sum of the statistical and systematic errors. The best-fit linear
expansion in radius of 4000\,km\,s$^{-1}$ is shown by the straight line.\label{plot}}
\end{figure}

\begin{figure}[!ht]
\epsscale{0.8}
\plotone{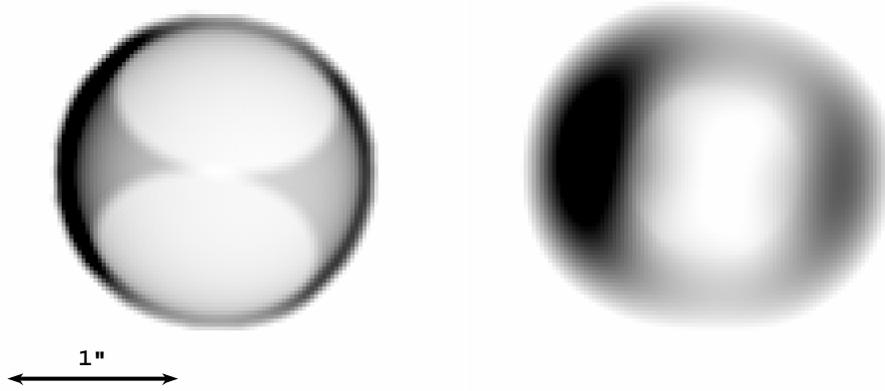}
\caption{\emph{Left:} The best-fit model for the most recent epoch (2008.0),
with the parameters listed in Table~\ref{tab2}.
\emph{Right:} The same model convolved with a 0\farcs4 FWHM Gaussian kernel,
as the super-resolved beam used in Fig.~\ref{super}. The brightness scales in
these images are arbitrary. \label{model}}
\end{figure}

\begin{figure}[!ht]
\epsscale{0.7}
\plotone{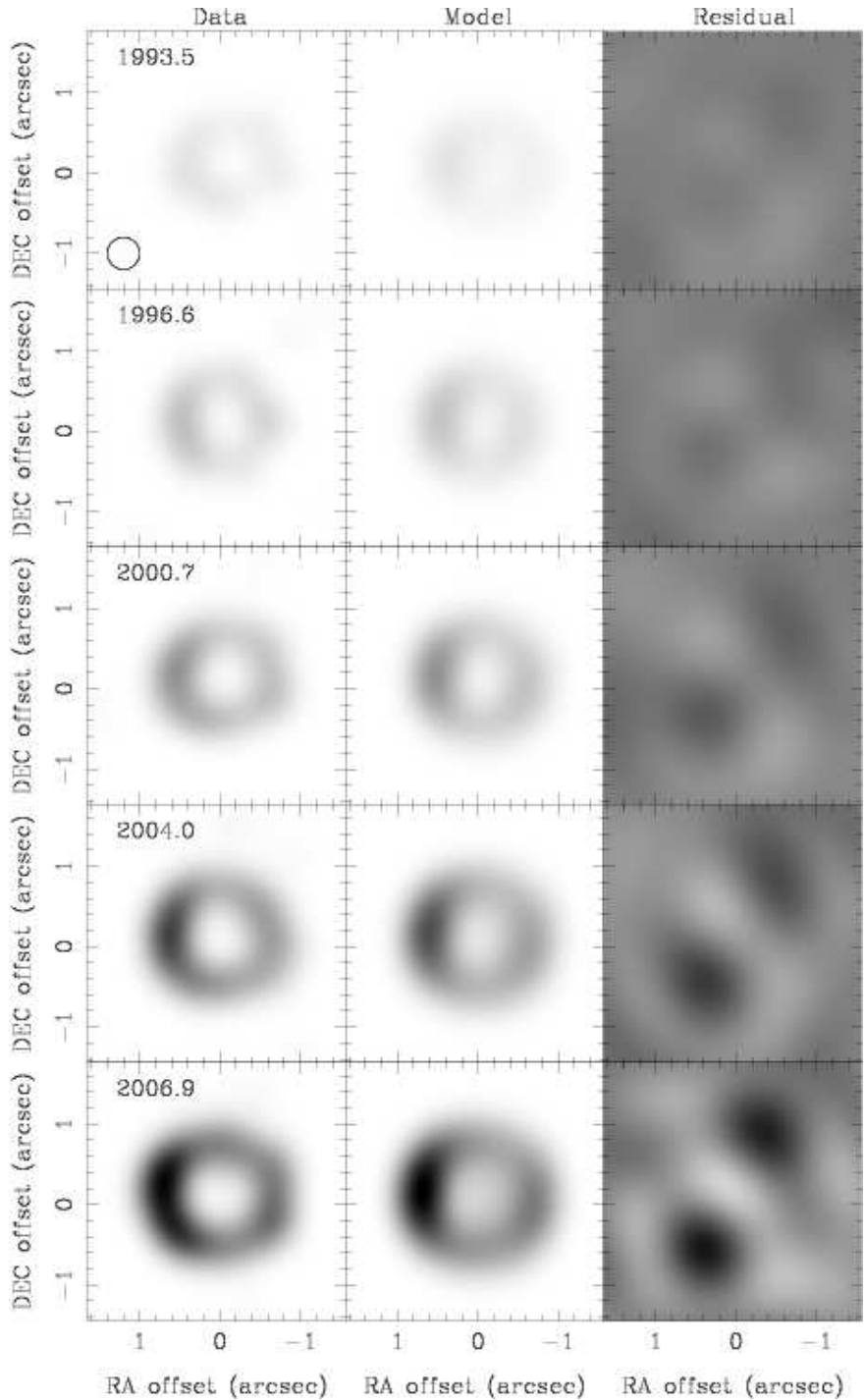}
\caption{Super-resolved images at 9\,GHz (\emph{left column}) compared to the
best-fit models (\emph{middle column}) for selected epochs. These images were
generated with the same procedure described in \S\ref{s3}. The synthesized beam,
which is a circular Gaussian with FWHM $0\farcs4$, is shown in the first panel.
The right column shows the dirty maps of the residual visbilities, for which no
deconvolution has been applied.
The scales are linear, ranging from 0 to 10\,mJy\,beam$^{-1}$
in the left and middle panels and from --1.5 to 1.5\,mJy\,beam$^{-1}$
for the residuals.\label{fits}}
\end{figure}

\begin{figure}[!ht]
\epsscale{0.95}
\plotone{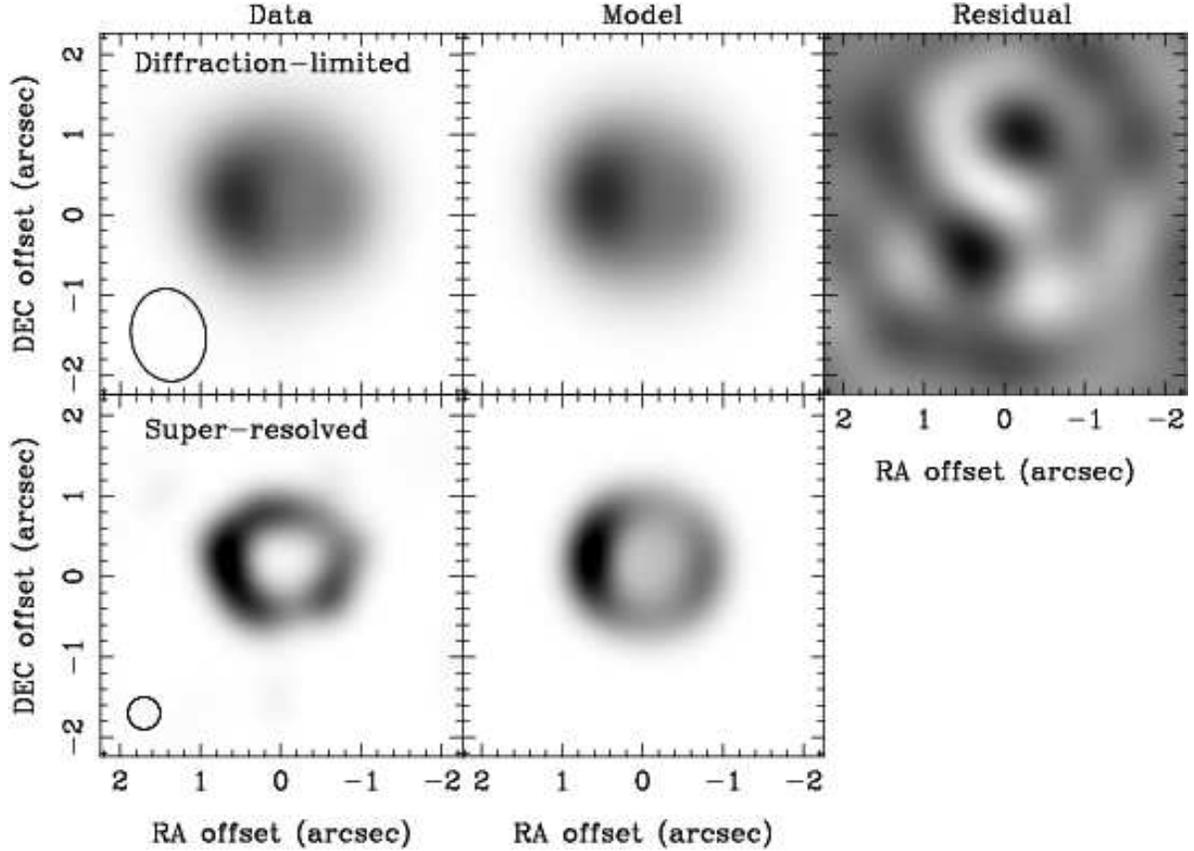}
\caption{Super-resolved images at 9\,GHz (\emph{left column}) compared to the
best-fit models (\emph{middle column}) for the last epoch (2008.0). The scales are
all linear, ranging from 10 to 40\,mJy\,beam$^{-1}$ for the diffraction-limited data and
model images, and from 0 to 10\,mJy\,beam$^{-1}$ for the super-resolved images. As
in Fig.~\ref{fits}, the dirty map of the residuals (\emph{right column}), which are
not deconvolved, has a linear scale ranging from from --1.5 to 1.5\,mJy\,beam$^{-1}$.
The synthesized beams are shown by the open ellipse and circle in the left column.
Note that the model images shown here have been generated with identical $u$-$v$
sampling as the data, while Fig.~\ref{model} was obtained by direct Fourier transform of
the model visibility assuming a complete $u$-$v$ coverage. \label{2008}}
\end{figure}

\begin{figure}[!ht]
\epsscale{0.95}
\plotone{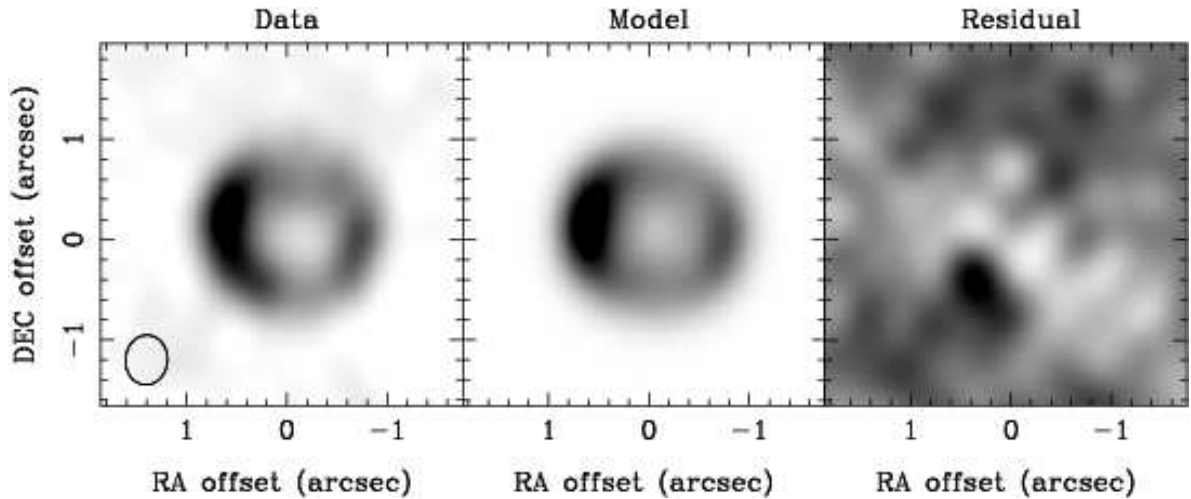}
\caption{18\,GHz diffraction-limited images of the data, model and the residual map
for epoch 2003.6 \citep{man05}. The synthesized beam is shown in the left panel.
The linear scales range from 0 to $3.5\;\mathrm{mJy\;beam^{-1}}$
in left two panels and and from $-0.5$ to $0.5\;\mathrm{mJy\;beam^{-1}}$
for the residual map. Note that no deconvolution has been applied to the latter.\label{12mm}}
\end{figure}

\begin{figure}[!ht]
\epsscale{1.0}
\plotone{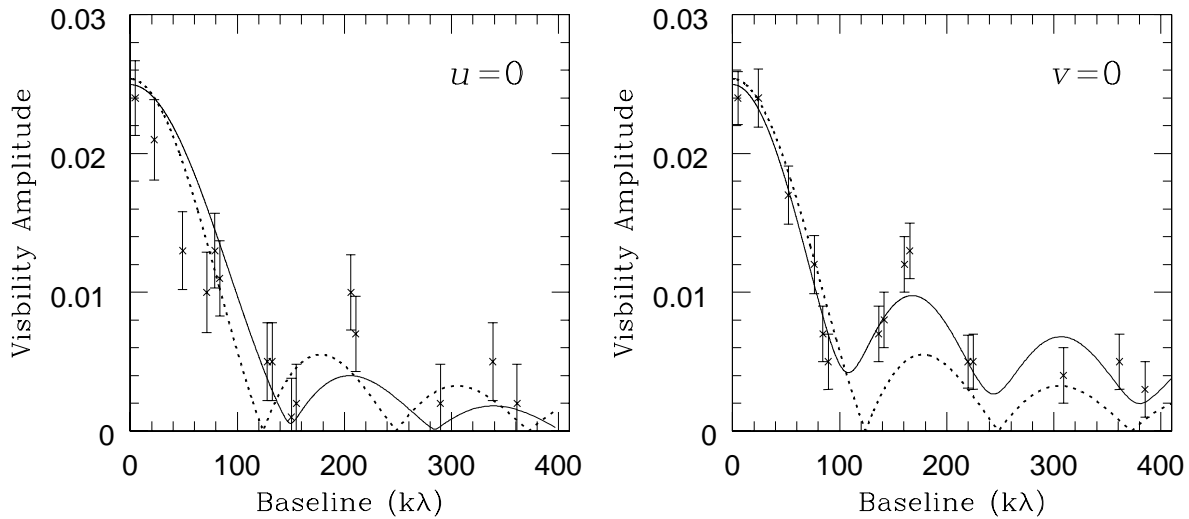}
\caption{Similar plot as Fig.~\ref{vis} for the 18\,GHz observation at epoch
2003.6. Note that the left panel is actually a slice at 10\arcdeg\ to the
$v$-axis, due to the $u$-$v$ coverage of the observation. \label{vis12}}
\end{figure}

\begin{figure}[!ht]
\epsscale{0.8}
\plotone{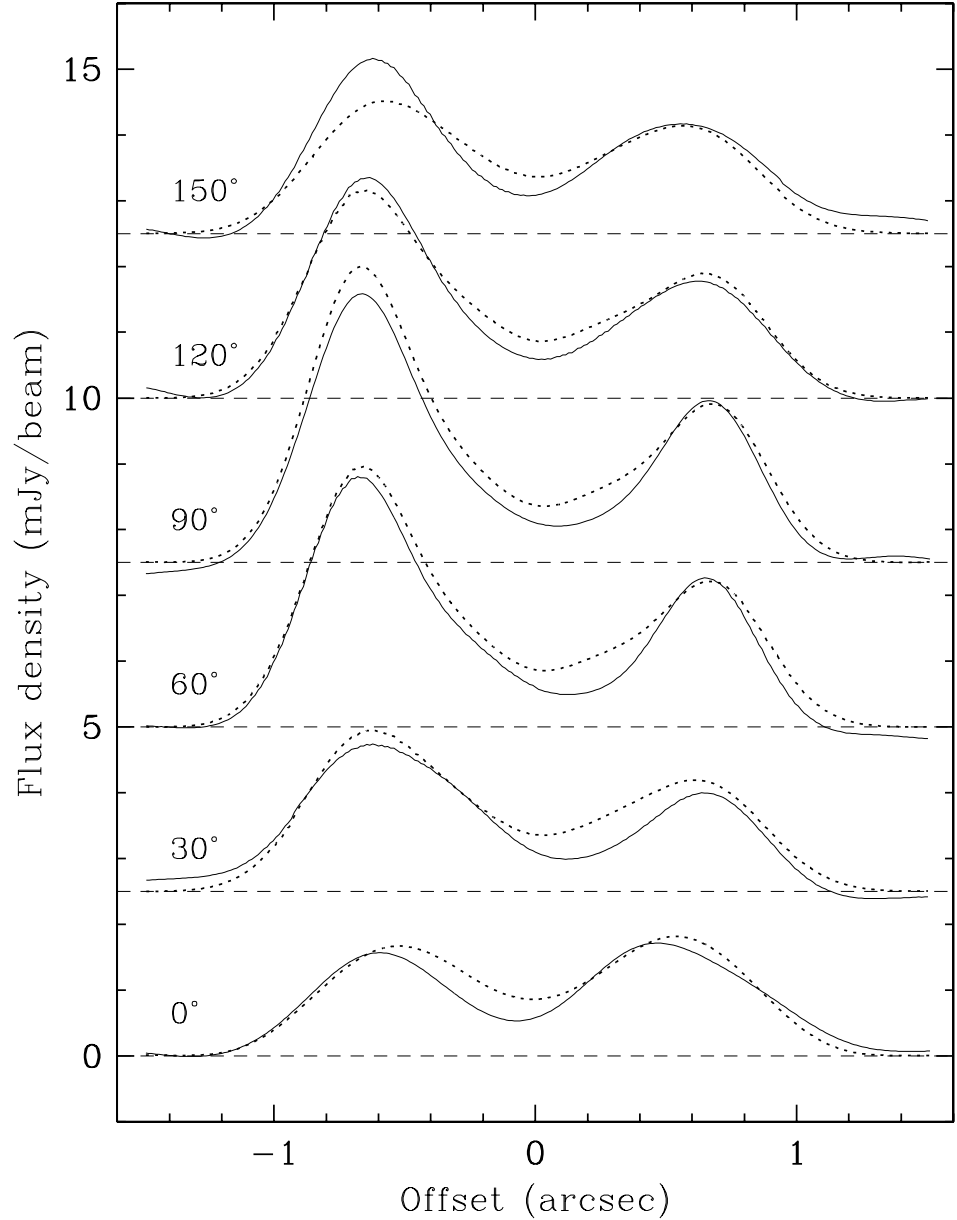}
\caption{Slices through the 18\,GHz diffraction-limited images in Fig.~\ref{12mm}
at different position angles. The offset is radial from the torus center and is positive
toward north and/or east. The data and model are shown by the solid and dotted lines
respectively. The zero levels for each slice are shown by the dashed lines.
\label{slice}}
\end{figure}

We first set up a coordinate system $(x',y',z')$ with $z'$ along the symmetric axis and
$y'$ along the projected line of sight at the equatorial plane. In this frame, a model
torus (or any axisymmetric object) can be built up from rings centered at $(0,z')$ with
radius $r$
\[ I^{\rm torus}(x',y') = \int \!\!\! \int w \ I^{\rm ring}_{r,z'} \;dr \; dz' \ , \]
where the intensity of each ring is given by
$I^{\rm ring}_{r,z'}=f\delta \left(r-\sqrt{x'^2+y'^2}\right)$.
In our case, the weight factor has value $w\equiv 1$ as the torus has a uniform emissivity.
Applying the Hankel Transform \citep{bra00}, each ring in the Fourier domain is given by
\[ F^{\rm ring}(u,v)\equiv\mathcal{F} \{I^{\rm ring}\} = J_0 \left
( 2\pi\,r \sqrt{u^2+v^2} \right) \]
where $J_0$ is the Bessel Function of the First Kind. In order to account for the asymmetry
observed in the remnant, we assume that the flux density of the torus varies linearly at a
rate $g$ along the direction $\phi$ in the equatorial plane (see Fig.~\ref{draw}). Hence the
intensity of each ring is modified to
\[ I^{\rm asym}=[g(x'\cos\phi + y'\sin\phi) +1] I^{\rm ring} \ . \]
With the Convolution Theorem \citep{bra00}, the Fourier transform of an asymmetric ring
is then
\begin{eqnarray*}
F^{\rm asym}(u,v) & \equiv & \mathcal{F} \{ I^{\rm asym} \} \\
& = & \mathcal{F} \{ g(x'\cos\phi + y'\sin\phi) +1 \} \ast\ast  \mathcal{F} \{I^{\rm ring}\} \\
& = & \left \{\frac{ig}{2\pi} \left [\cos\phi \delta'(u)\delta(v) + \sin\phi \delta(u)\delta'(v) \right ] + \delta(u)\delta(v) \right \} \ast\ast F^{\rm ring}(u,v) \\
& = & \left [\frac{ig}{2\pi}\left(\cos\phi\frac{\partial}{\partial u} +
\sin\phi\frac{\partial}{\partial v}\right ) +1\right] F^{\rm ring}(u,v) \\
&= & -\frac{irg}{\sqrt{u^2+v^2}}(u\sin\phi+v\cos\phi)J_1 \left ( 2\pi\,r \sqrt{u^2+v^2}
  \right) + J_0 \left ( 2\pi\,r \sqrt{u^2+v^2} \right) \ ,
\end{eqnarray*}
where $\ast \ast$ denotes a 2D convolution such that
\[ F**G \equiv \int_{-\infty}^{+\infty} \int_{-\infty}^{+\infty} F(u,v) G(u-u',v-v')
\,du'\,dv' \] for any 2D functions $F(u,v)$ and $G(u,v)$. $\delta (u), \delta(v)$ are Dirac's $\delta$ functions and $\delta'(u)\equiv \case{d\delta}{du}$ and $\delta'(v)\equiv \case{d\delta}{dv}$.

The model torus has to be projected onto the plane of the sky before it is compared to
the observations. Here we set up another coordinate system $(x,y,z)$, with $z$ along
the observer line of sight and $y$ along the projected torus axis on the image plane.
The transformation between the two frames is given by
\[ \left \{ \begin{array}{cl}
x & = x' \\
y & = y'\sin\zeta-z'\cos\zeta \\
z & = y'\cos\zeta+z'\sin\zeta \ , \end{array} \right . \]
where $\zeta$ is the inclination angle between the torus equatorial plane and the observer
line of sight (see Fig.~\ref{draw}). The projection essentially compresses the rings into
ellipses in the image plane, with center at $(x,y)=(0,-z'\cos \zeta)$ and semi-minor axis
$r\sin \zeta$. Hence an asymmetric ring projected on the image plane can be expressed as
\[ I^{\rm proj}_{r,z'} (x,y)= \left [ g \left( x\cos\phi + \case{y+z'\cos\zeta}{r\sin\zeta} 
\sin\phi  \right ) + 1 \right ] f\delta\left(r-\sqrt{x^2+\left(
\case{y+z'\cos\zeta}{\sin\zeta} \right )^2}\right) \ . \]
Following the scaling and shifting properties of the Fourier Transform \citep{bra00},
for any constants $a$ and $b$,
\begin{eqnarray*}
\mathcal{F} \{ f(x-a,y-b) \}& = & e^{-i2\pi(au+bv)} \mathcal{F} \{ f \} (u,v) \\
\mathcal{F} \{ f(ax,by) \}& = & \frac{1}{|ab|}\mathcal{F} \{ f \}  \left(\frac{u}{a},\frac{v}{b}\right) \ ,
\end{eqnarray*}
hence \[ F^{\rm proj}(u,v)\equiv \mathcal{F} \{I^{\rm proj}\}=
e^{i2\pi z'v\cos\zeta}F^{\rm asym}(u,v\sin\zeta) \]
up to an overall normalization factor.

To correct for light travel time effects, we assume the torus flux density $f$
increases linearly during the light travel time across the remnant, i.e.\
$f\approx \dot {f}t+f_0 $, for constants $\dot {f}$ and $f_0$. While the latter
is a fitting parameter, the former is determined from the local derivative
of the light curve given by \citet{sta07}. The relative light traveling time
from each point of the ring to the observer is 
\[ \Delta t = \frac{z}{c} = \frac{d}{c} \left(y'\cos\zeta+z'\sin\zeta \right)
=\frac{d}{c}\left(\frac{y}{\tan \zeta}+\frac{z'}{\sin\zeta} \right) \ , \]
where $d=51\,$kpc is the distance to LMC, and $z$ is in the units of radians.
Hence the flux density at each point is
\[ f=\frac{d\dot{f}}{c}\left(\frac{y}{\tan \zeta}+\frac{z'}
{\sin\zeta} \right) + f_0  \] 
and the torus intensity is modified by a linear gradient of
\[ \frac{d}{c}\frac{\dot{f}}{f}\left(\frac{y}{\tan \zeta}+\frac{z'}
{\sin\zeta} \right) +1\ . \]
Employing the Convolution Theorem again, we obtain the Fourier transform
of an asymmetric ring including a light travel time correction
\[ F^{\rm light}(u,v)=\left[ \frac{d}{c}\frac{\dot{f}}{f} \left(
\frac{i}{2\pi}\frac{1}{\tan\zeta}\frac{\partial}{\partial v}+
\frac{z'}{\sin\zeta}\right)+1\right] F^{\rm proj}(u,v) \ . \] 

Finally, the model is shifted and rotated according to $(x_0,y_0)$ and $\Psi$
respectively, \[ F^{\rm final}(u,v)= e^{-i2\pi(x_{0}u+y_{0}v)}F^{\rm light}
(u\cos\Psi-v\sin\Psi,u\sin\Psi+v\cos\Psi) \]
and integrated through $r$ and $z'$ since the Fourier transform is linear.
\[ F^{\rm torus}(u,v)=\int \!\!\! \int w F^{\rm final}_{r,z'} \;dr \; dz' \ . \]

\clearpage

\clearpage 
\begin{deluxetable}{lcclcc}
\tablecaption{Observational parameters for the datasets used in this study.\label{tab1}}
\tablewidth{0pt}
\tabletypesize{\scriptsize}
\tablehead{
\colhead{Observing date} & \colhead{Days since} & \colhead{Array} &
\colhead{Frequency} & \colhead{Time on} & \colhead{Epoch shown}  \\
\colhead{} & \colhead{supernova} & \colhead{Configuration} &
\colhead{(MHz)} & \colhead{source (hr)} & \colhead{in Fig.~\ref{natural}
\& \ref{super}\tablenotemark{\dag}}}
\tablecolumns{5}
\startdata
\sidehead{9\,GHz Observations}
1992 Jan 14& 1786 & 6B & 8640  & 12 & \nodata \\
1992 Mar 20& 1852 & 6A & 8640  & 10 & \nodata \\
1992 Oct 21& 2067 & 6C & 8640, 8900 & 13 & 1992.9 \\
1993 Jan 4 & 2142 & 6A & 8640, 8900 & 9 & 1992.9 \\
1993 Jan 5 & 2143 & 6A & 8640, 8900 & 6 & 1992.9 \\
1993 Jun 24& 2313 & 6C & 8640, 8900 & 8 & 1993.6 \\
1993 Jul 1 & 2320 & 6C & 8640, 8900 & 10 & 1993.6 \\
1993 Oct 15& 2426 & 6A & 8640, 9024 & 17 & 1993.6 \\
1994 Feb 16& 2550 & 6B & 8640, 9024 & 9 & 1994.4 \\
1994 Jun 27-28& 2681 & 6C & 8640, 9024 & 21 & 1994.4 \\
1994 Jul 1 & 2685 & 6A & 8640, 9024 & 10 & 1994.4 \\
1995 Jul 24& 3073 & 6C & 8640, 9024 & 12 & 1995.7 \\
1995 Aug 29& 3109 & 6D & 8896, 9152 & 7 & 1995.7 \\
1995 Nov 6 & 3178 & 6A & 8640, 9024 & 9 & 1995.7 \\
1996 Jul 21& 3436 & 6C & 8640, 9024 & 14 & 1996.7 \\
1996 Sep 8 & 3485 & 6B & 8640, 9024 & 13 & 1996.7 \\
1996 Oct 5 & 3512 & 6A & 8896, 9152 & 8  & 1996.7 \\
1997 Nov 11& 3914 & 6C & 8512, 8896 & 7  & 1998.0 \\
1998 Feb 18& 4013 & 6A & 8896, 9152 & 10 & 1998.0  \\
1998 Feb 21& 4016 & 6B & 8512, 9024 & 7  & 1998.0 \\
1998 Sep 13& 4220 & 6A & 8896, 9152 & 12 & 1998.9  \\
1998 Oct 31& 4268 & 6D & 8502, 9024 & 11 & 1998.9  \\
1999 Feb 12& 4372 & 6C & 8512, 8896 & 10 & 1999.7 \\
1999 Sep 5 & 4577 & 6D & 8768, 9152 & 11 & 1999.7 \\
1999 Sep 12& 4584 & 6A & 8512, 8896 & 14 & 1999.7 \\
2000 Sep 28& 4966 & 6A & 8512, 8896 & 10 & 2000.9 \\
2000 Nov 12& 5011 & 6C & 8512, 8896 & 11 & 2000.9 \\
2001 Nov 23& 5387 & 6D & 8768, 9152 & 8 & 2001.9 \\
2002 Nov 19& 5748 & 6A & 8512, 8896 & 8 & 2003.0 \\
2003 Jan 20& 5810 & 6B & 8512, 9024 & 9 & 2003.0 \\
2003 Aug 1 & 6003 & 6D & 8768, 9152 & 10 & 2003.6 \\
2003 Dec 5 & 6129 & 6A & 8512, 8896 & 9 & 2004.0 \\
2004 Jan 15& 6170 & 6A & 8512, 8896 & 9 & 2004.0 \\
2004 May 7 & 6283 & 6C & 8512, 8896 & 9 & 2004.4 \\
2005 Mar 25& 6605 & 6A & 8512, 8896 & 9 & 2005.2 \\
2005 Jun 21& 6693 & 6B & 8512, 8896 & 9 & 2005.5 \\
2006 Mar 28& 6973 & 6C & 8512, 8896 & 9 & 2006.2 \\
2006 Jul 18& 7085 & 6A & 8512, 8896 & 9 & 2006.5 \\
2006 Dec 8 & 7228 & 6B & 8512, 9024 & 8 & 2006.9 \\
2008 Jan 4 & 7620 & 6A & 8512, 9024 & 11 & 2008.0 \\ 
\sidehead{18\,GHz Observations}
2003 Jul 31& 6002 & 6D & 17345, 19649 & 8 & \nodata \\
2004 May 6 & 6282 & 6C & 17345, 19649 & 7 & \nodata
\enddata
\tablenotetext{\dag}{Some datasets have been averaged together
to generate the corresponding images in Fig.~\ref{natural} and
\ref{super} for the listed epoch.}
\end{deluxetable}

\begin{deluxetable}{ccccccccc}
\tablecaption{Best-fit torus parameters for each epoch, with 1$\sigma$ 
confidence intervals.\label{tab2}}
\tablewidth{0pt}
\tabletypesize{\scriptsize}
\tablehead{
\colhead{Day}&\colhead{$f$ (mJy)}&\colhead{$R$ (\arcsec)}&
\colhead{$\theta$ (\arcdeg)}&\colhead{$\delta$ (\%)}&
\colhead{$\phi$ (\arcdeg)}&\colhead{$g$ (\%)}&\colhead{$\chi^2/\nu$\tablenotemark{\dag}}
&\colhead{dof\tablenotemark{\S}}}
\startdata
\sidehead{9\,GHz Observations}
1786 & $3.9^{+0.2}_{-0.2}$ & $0.60^{+0.07}_{-0.04}$ & $70^{+10}_{-14 }$ & $162^{+19}_{-49}$ & $188^{+20}_{-12}$ & $83^{+16}_{-26}$ & 1.8&2115 \\
1852 & $3.7^{+0.2}_{-0.2}$ & $0.63^{+0.04}_{-0.08}$ & $46^{+19}_{-40 }$ & $108^{+66}_{-62}$ & $57^{+54}_{-19}$ & $76^{+24}_{-25}$ & 3.7&1650 \\
2067 & $5.5^{+0.1}_{-0.2}$ & $0.73^{+0.01}_{-0.07}$ & $56^{+2}_{-18 }$ & $108^{+72}_{-14}$ & $143^{+3}_{-23}$ & $99^{+1}_{-19}$ & 4.3&3610 \\
2142 & $5.0^{+0.2}_{-0.2}$ & $0.66^{+0.03}_{-0.02}$ & $63^{+3}_{-18 }$ & $81^{+101}_{-45}$ & $142^{+15}_{-35}$ & $45^{+54}_{-16}$ & 17&2710 \\
2143 & $5.6^{+0.2}_{-0.1}$ & $0.64^{+0.02}_{-0.01}$ & $0.6^{+24}_{-0.4 }$ & $1^{+37}_{-1}$ & $92^{+25}_{-15}$ & $39^{+13}_{-5}$ & 16&1400 \\
2313 & $6.7^{+0.1}_{-0.1}$ & $0.62^{+0.01}_{-0.01}$ & $33^{+5}_{-8 }$ & $1^{+33}_{-1}$ & $83^{+9}_{-11}$ & $40^{+7}_{-3}$ & 3.6&2910 \\
2320 & $7.0^{+0.2}_{-0.2}$ & $0.67^{+0.02}_{-0.03}$ & $31^{+8}_{-15 }$ & $63^{+33}_{-48}$ & $78^{+12}_{-8}$ & $41^{+9}_{-7}$ & 4.5&2970 \\
2426 & $6.7^{+0.1}_{-0.1}$ & $0.690^{+0.009}_{-0.008}$ & $53^{+2}_{-2 }$ & $36^{+19}_{-35}$ & $75^{+9}_{-8}$ & $43^{+7}_{-5}$ & 5.2&4380 \\
2550 & $6.5^{+0.2}_{-0.2}$ & $0.73^{+0.02}_{-0.06}$ & $57^{+3}_{-25 }$ & $91^{+85}_{-23}$ & $133^{+9}_{-27}$ & $81^{+15}_{-15}$ & 7.5&3000 \\
2681 & $8.4^{+0.1}_{-0.1}$ & $0.672^{+0.005}_{-0.005}$ & $46^{+1}_{-2 }$ & $23^{+17}_{-23}$ & $78^{+6}_{-7}$ & $35^{+4}_{-2}$ & 6.0&6150 \\
2685 & $8.1^{+0.1}_{-0.1}$ & $0.665^{+0.007}_{-0.007}$ & $52^{+2}_{-2 }$ & $1^{+16}_{-1}$ & $106^{+8}_{-10}$ & $37^{+3}_{-2}$ & 5.7&3264 \\
3073 & $11.1^{+0.2}_{-0.2}$ & $0.671^{+0.008}_{-0.008}$ & $32^{+3}_{-4 }$ & $52^{+11}_{-13}$ & $80^{+5}_{-4}$ & $41^{+2}_{-2}$ & 5.6&2670 \\
3109 & $9.7^{+0.4}_{-0.3}$ & $0.633^{+0.052}_{-0.02}$ & $16^{+26}_{-15 }$ & $1^{+17}_{-0.5}$ & $75^{+12}_{-6}$ & $43^{+5}_{-3}$ & 17&1606 \\
3178 & $11.7^{+0.1}_{-0.1}$ & $0.685^{+0.005}_{-0.005}$ & $44^{+2}_{-2 }$ & $0.4^{+3}_{-0.3}$ & $94^{+4}_{-4}$ & $38.0^{+1.4}_{-1.2}$ & 4.0&3450 \\
3436 & $15.2^{+0.1}_{-0.1}$ & $0.705^{+0.004}_{-0.003}$ & $46.3^{+0.9}_{-1.0 }$ & $26^{+8}_{-14}$ & $95^{+3}_{-4}$ & $41.9^{+1.4}_{-1.2}$ & 4.9&4345 \\
3485 & $15.4^{+0.1}_{-0.1}$ & $0.707^{+0.004}_{-0.004}$ & $50.1^{+1.0}_{-1.0 }$ & $0.4^{+7}_{-0.29}$ & $102^{+3}_{-3}$ & $43.3^{+1.4}_{-1.3}$ & 5.2&4710 \\
3512 & $15.4^{+0.1}_{-0.1}$ & $0.708^{+0.005}_{-0.006}$ & $51.1^{+1.2}_{-1.4 }$ & $0.1^{+25}_{-0.1}$ & $94^{+5}_{-5}$ & $42^{+3}_{-2}$ & 3.3&2640 \\
3914 & $17.6^{+0.2}_{-0.2}$ & $0.694^{+0.007}_{-0.006}$ & $41^{+2}_{-2 }$ & $0.04^{+0.5}_{-0.04}$ & $112^{+4}_{-5}$ & $39^{+2}_{-2}$ & 2.4&1280 \\
4013 & $19.1^{+0.1}_{-0.1}$ & $0.754^{+0.004}_{-0.004}$ & $45.5^{+1.0}_{-1.0 }$ & $0.3^{+0.3}_{-0.3}$ & $105^{+2}_{-2}$ & $45.1^{+1.2}_{-1.0}$ & 3.0&2838 \\
4016 & $18.7^{+0.1}_{-0.1}$ & $0.745^{+0.005}_{-0.004}$ & $49.4^{+1.1}_{-1.1 }$ & $0.1^{+0.1}_{-0.1}$ & $103^{+3}_{-3}$ & $45.7^{+1.1}_{-1.0}$ & 3.1&2520 \\
4220 & $20.2^{+0.1}_{-0.1}$ & $0.729^{+0.003}_{-0.004}$ & $42.6^{+0.9}_{-1.0 }$ & $6^{+8}_{-5}$ & $100^{+2}_{-2}$ & $37.6^{+0.8}_{-0.8}$ & 2.2&2963 \\
4268 & $21.8^{+0.3}_{-0.3}$ & $0.736^{+0.004}_{-0.004}$ & $40.0^{+1.0}_{-1.0 }$ & $30^{+8}_{-15}$ & $107^{+2}_{-2}$ & $38.0^{+0.9}_{-0.8}$ & 7.3&3870 \\
4372 & $22.9^{+0.2}_{-0.2}$ & $0.727^{+0.004}_{-0.004}$ & $36.6^{+1.1}_{-1.2 }$ & $24^{+6}_{-10}$ & $114^{+2}_{-2}$ & $37.5^{+0.8}_{-0.7}$ & 3.6&3540 \\
4577 & $23.9^{+0.3}_{-0.2}$ & $0.757^{+0.005}_{-0.005}$ & $39.9^{+0.8}_{-1.1 }$ & $22^{+9}_{-19}$ & $103^{+2}_{-2}$ & $39.4^{+0.9}_{-0.9}$ & 7.0&3450 \\
4584 & $25.2^{+0.09}_{-0.08}$ & $0.747^{+0.002}_{-0.002}$ & $41.4^{+0.7}_{-0.7 }$ & $0.4^{+0.1}_{-0.3}$ & $109^{+2}_{-2}$ & $38.5^{+0.6}_{-0.6}$ & 2.9&4230 \\
4966 & $29.4^{+0.05}_{-0.06}$ & $0.764^{+0.002}_{-0.002}$ & $40.0^{+0.6}_{-0.7 }$ & $0.3^{+0.3}_{-0.2}$ & $108.5^{+1.3}_{-1.3}$ & $39.6^{+0.5}_{-0.5}$ & 1.3&50697 \\
5011 & $33.0^{+0.07}_{-0.08}$ & $0.776^{+0.002}_{-0.002}$ & $43.4^{+0.5}_{-0.7 }$ & $0.1^{+0.7}_{-0.02}$ & $104.9^{+1.4}_{-1.3}$ & $40.1^{+0.6}_{-0.5}$ & 1.4&50539 \\
5387 & $34.1^{+0.1}_{-0.1}$ & $0.789^{+0.003}_{-0.003}$ & $40.7^{+1.0}_{-0.9 }$ & $0.01^{+3.4}_{-0.01}$ & $108.2^{+1.1}_{-1.4}$ & $41.4^{+0.5}_{-0.6}$ & 1.6&39612 \\
5748 & $41.7^{+0.06}_{-0.08}$ & $0.811^{+0.001}_{-0.002}$ & $43.0^{+0.5}_{-0.5 }$ & $0.10^{+3.4}_{-0.05}$ & $103.0^{+1.2}_{-0.9}$ & $40.4^{+0.3}_{-0.5}$ & 1.3&39000 \\
5810 & $42.5^{+0.07}_{-0.07}$ & $0.815^{+0.001}_{-0.001}$ & $42.2^{+0.5}_{-0.4 }$ & $0.001^{+0.4}_{-0.001}$ & $118.1^{+0.9}_{-0.9}$ & $43.2^{+0.4}_{-0.5}$ & 1.6&46020 \\
6003 & $46.5^{+0.1}_{-0.08}$ & $0.813^{+0.003}_{-0.001}$ & $38.8^{+1.0}_{-0.4 }$ & $0.06^{+3.7}_{-0.01}$ & $100.3^{+2}_{-0.04}$ & $38.2^{+0.5}_{-0.3}$ & 1.6&45000 \\
6129 & $52.8^{+0.08}_{-0.07}$ & $0.833^{+0.002}_{-0.002}$ & $41.9^{+0.6}_{-0.4 }$ & $0.01^{+0.3}_{-0.01}$ & $107.7^{+0.9}_{-1.0}$ & $42.2^{+0.4}_{-0.4}$ & 1.3&46020 \\
6170 & $54.0^{+0.1}_{-0.07}$ & $0.831^{+0.002}_{-0.001}$ & $41.4^{+0.6}_{-0.4 }$ & $0.17^{+2.4}_{-0.14}$ & $108.1^{+1.3}_{-0.6}$ & $38.8^{+0.5}_{-0.3}$ & 1.5&43680 \\
6283 & $53.6^{+0.08}_{-0.06}$ & $0.829^{+0.001}_{-0.001}$ & $39.0^{+0.5}_{-0.5 }$ & $0.16^{+1.0}_{-0.11}$ & $107.3^{+1.0}_{-0.6}$ & $38.7^{+0.4}_{-0.3}$ & 1.5&44850 \\
6605 & $61.3^{+0.1}_{-0.1}$ & $0.844^{+0.002}_{-0.002}$ & $37.7^{+0.5}_{-0.6 }$ & $0.19^{+2.1}_{-0.14}$ & $108.2^{+1.0}_{-0.6}$ & $39.0^{+0.3}_{-0.3}$ & 2.9&42900 \\
6693 & $62.7^{+0.07}_{-0.08}$ & $0.858^{+0.001}_{-0.001}$ & $42.2^{+0.4}_{-0.4 }$ & $0.11^{+0.8}_{-0.06}$ & $100.6^{+0.9}_{-0.9}$ & $35.8^{+0.3}_{-0.3}$ & 1.5&39000 \\
6973 & $73.7^{+0.08}_{-0.07}$ & $0.881^{+0.001}_{-0.001}$ & $43.8^{+0.4}_{-0.3 }$ & $0.01^{+3.6}_{-0.01}$ & $116.4^{+0.6}_{-0.7}$ & $41.8^{+0.4}_{-0.3}$ & 1.5&40365 \\
7085 & $77.1^{+0.07}_{-0.09}$ & $0.872^{+0.001}_{-0.002}$ & $38.8^{+0.4}_{-0.5 }$ & $0.03^{+0.9}_{-0.001}$ & $110.5^{+0.5}_{-0.6}$ & $39.5^{+0.3}_{-0.3}$ & 1.5&29120 \\
7228 & $82.4^{+0.10}_{-0.09}$ & $0.874^{+0.002}_{-0.001}$ & $39.4^{+0.8}_{-0.4 }$ & $0.07^{+6.8}_{-0.04}$ & $107.3^{+0.5}_{-0.6}$ & $39.1^{+0.3}_{-0.2}$ & 1.4&35685 \\
7620 & $93.5^{+0.10}_{-0.10}$ & $0.893^{+0.001}_{-0.001}$ & $41.8^{+0.7}_{-0.4 }$ & $0.03^{+6.1}_{-0.00}$ & $103.7^{+0.5}_{-0.5}$ & $38.6^{+0.3}_{-0.3}$ & 1.3&42900 \\
\sidehead{18\,GHz Observations}
6002 & $25.1^{+0.08}_{-0.07}$ & $0.80^{+0.002}_{-0.002}$ & $32.1^{+0.6}_{-0.5 }$ & $0.03^{+0.3}_{-0.03}$ & $99.1^{+1.4}_{-1.4}$ & $36.5^{+0.6}_{-0.6}$ & 0.7&42939 \\
6282 & $24.3^{+0.10}_{-0.09}$ & $0.81^{+0.002}_{-0.002}$ & $26.9^{+0.6}_{-0.8 }$ & $10^{+6}_{-9}$ & $108.2^{+1.5}_{-1.4}$ & $36.2^{+0.9}_{-0.8}$ & 1.0&31200
\enddata
\tablenotetext{\dag}{The $\chi^2$ values include the statistical errors only, while the
uncertainties in the parameters include the systematic errors (see text).}
\tablenotetext{\S}{Before 2000, all 13 frequency channels in the data were averaged to boost the signal and to reduce the computing time.}
\end{deluxetable}

\begin{deluxetable}{cccc@{}l@{}|cccccccccccc}
\addtolength{\tabcolsep}{-3pt}
\tablecaption{Best-fit parameters for the shell fit and shell+2 points fit.
\label{tab3}}
\tablewidth{0pt}
\tabletypesize{\tiny}
\tablehead{
\colhead{}& \multicolumn{3}{c}{Fit to shell alone} & \colhead{}&
\multicolumn{12}{c}{Fit to shell + 2 points} \\
\cline{2-4} \cline{6-17} \\
\colhead{Day} & \colhead{$f_{\rm shell}$}&\colhead{$R$}&
\colhead{$\chi^2/\nu$}& \colhead{}&
\colhead{$f_{\rm shell}$}&\colhead{$R$}&
\colhead{$f_1$}&\colhead{$\delta{\rm RA}_1$}&\colhead{$\delta{\rm Dec}_1$}&
\colhead{$f_2$}&\colhead{$\delta{\rm RA}_2$}&\colhead{$\delta{\rm Dec}_2$}&
\colhead{$f_{\rm ratio}$\tablenotemark{\dag}} &
\colhead{$d_1/R$\tablenotemark{\S}} &
\colhead{$d_2/R$\tablenotemark{\S}} &
\colhead{$\chi^2/\nu$} \\
\colhead{} & \colhead{(mJy)} & \colhead{(\arcsec)} &
\colhead{} & \colhead{}&
\colhead{(mJy)} & \colhead{(\arcsec)} & 
\colhead{(mJy)} & \colhead{(\arcsec)} & \colhead{(\arcsec)} & 
\colhead{(mJy)} & \colhead{(\arcsec)} & \colhead{(\arcsec)} }
\startdata
1786&3.8&0.67&1.8&&2.5&0.97&1.2&-0.04&-0.09&0.4&-0.83&-0.05&1.5&0.10&0.86&1.8\\
1852&3.6&0.64&3.7&&2.6&0.57&0.5&-0.02&-0.57&0.5&-0.88&-0.02&1.0&1.00&1.54&3.7\\
2067&5.3&0.67&4.4&&4.0&0.81&1.1&0.17&-0.03&0.3&-0.71&-0.34&1.3&0.21&0.98&4.3\\
2142&4.9&0.67&17&&4.1&0.76&0.7&0.17&-0.05&0.2&-0.44&-0.04&1.2&0.23&0.58&17\\
2143&5.6&0.63&16&&4.2&0.67&1.1&0.41&-0.06&0.3&-0.76&-0.10&1.3&0.61&1.15&16\\
2313&6.7&0.62&3.6&&5.1&0.64&1.2&0.36&-0.14&0.5&-0.74&-0.04&1.2&0.61&1.15&3.6\\
2320&7.0&0.67&4.5&&5.3&0.72&1.2&0.36&-0.07&0.5&-0.69&0.14&1.2&0.51&0.97&4.5\\
2426&6.6&0.67&5.3&&5.3&0.76&1.1&0.31&-0.07&0.3&-0.48&0.10&1.3&0.42&0.64&5.2\\
2550&6.3&0.71&7.5&&4.9&0.76&0.6&0.77&0.61&0.9&0.29&0.03&0.9&1.29&0.39&7.5\\
2681&8.4&0.66&6.1&&6.7&0.72&1.3&0.30&0.00&0.5&-0.70&0.11&1.2&0.41&0.98&6.0\\
2685&8.0&0.66&5.7&&6.5&0.70&1.2&0.30&-0.07&0.4&-0.78&0.11&1.2&0.44&1.12&5.6\\
3073&11&0.69&5.9&&8.8&0.70&1.8&0.45&-0.06&0.5&-0.86&-0.10&1.3&0.66&1.24&5.6\\
3109&10&0.69&17&&6.7&0.66&2.2&0.43&-0.03&0.7&-0.59&0.07&1.4&0.66&0.90&17\\
3178&12&0.69&4.2&&9.2&0.73&2.0&0.39&-0.07&0.5&-0.65&-0.04&1.3&0.55&0.90&4.0\\
3436&15&0.69&5.3&&12&0.76&2.5&0.35&-0.04&0.6&-0.72&0.08&1.3&0.46&0.95&4.9\\
3485&15&0.69&5.4&&13&0.76&2.5&0.36&-0.01&0.3&-0.64&0.03&1.3&0.47&0.84&5.1\\
3512&15&0.69&3.5&&12&0.77&2.6&0.37&-0.03&0.4&-0.58&0.15&1.3&0.48&0.77&3.3\\
3914&17&0.72&2.8&&14&0.76&2.9&0.43&-0.08&0.7&-0.82&0.32&1.3&0.57&1.15&2.4\\
4013&19&0.74&3.6&&15&0.80&3.2&0.41&0.00&0.4&-0.70&-0.09&1.3&0.51&0.88&3.0\\
4016&19&0.74&3.6&&14&0.77&3.5&0.41&0.01&0.7&-0.62&0.23&1.4&0.54&0.85&3.0\\
4220&20&0.73&2.7&&16&0.77&3.2&0.41&-0.02&0.9&-0.79&0.05&1.3&0.54&1.03&2.2\\
4268&22&0.73&7.9&&17&0.80&3.6&0.38&-0.02&1.0&-0.77&-0.07&1.3&0.48&0.97&7.3\\
4372&23&0.75&4.3&&18&0.78&3.6&0.40&-0.04&1.3&-0.83&-0.08&1.2&0.52&1.07&3.6\\
4577&24&0.76&7.8&&19&0.81&4.0&0.42&0.00&1.0&-0.78&0.02&1.3&0.51&0.96&7.0\\
4584&25&0.74&3.6&&20&0.79&4.0&0.43&-0.03&1.1&-0.80&-0.07&1.3&0.55&1.02&2.8\\
4966&29&0.76&1.4&&23&0.81&4.8&0.43&-0.03&1.4&-0.83&-0.07&1.3&0.54&1.03&1.3\\
5011&33&0.77&1.5&&27&0.82&5.2&0.43&-0.03&1.2&-0.81&-0.04&1.3&0.53&0.99&1.4\\
5387&34&0.80&1.7&&27&0.83&5.5&0.45&-0.03&1.4&-0.83&-0.08&1.3&0.54&1.00&1.6\\
5748&41&0.81&1.5&&34&0.85&6.5&0.46&-0.02&1.6&-0.87&-0.01&1.3&0.55&1.02&1.3\\
5810&42&0.82&1.7&&34&0.85&6.4&0.47&0.01&1.6&-0.92&-0.14&1.3&0.55&1.09&1.6\\
6003&46&0.83&1.8&&38&0.85&7.1&0.50&-0.03&1.8&-0.88&-0.05&1.3&0.59&1.03&1.6\\
6129&53&0.84&1.5&&42&0.88&8.6&0.47&-0.02&1.8&-0.85&-0.07&1.3&0.54&0.97&1.3\\
6170&54&0.85&1.7&&44&0.87&8.4&0.50&-0.02&2.0&-0.84&-0.08&1.3&0.57&0.97&1.5\\
6283&53&0.84&1.8&&44&0.87&7.8&0.50&-0.01&1.9&-0.96&-0.09&1.2&0.58&1.11&1.5\\
6605&61&0.86&3.3&&49&0.88&9.5&0.51&0.00&2.6&-0.91&-0.07&1.3&0.58&1.04&2.9\\
6693&62&0.87&1.8&&52&0.88&8.6&0.54&-0.01&2.5&-1.00&-0.03&1.2&0.61&1.13&1.5\\
6973&73&0.90&1.9&&60&0.92&10&0.48&0.05&2.8&-0.94&-0.07&1.2&0.53&1.02&1.5\\
7085&77&0.89&2.3&&64&0.92&11&0.52&0.02&2.7&-0.92&-0.11&1.2&0.57&1.01&1.6\\
7228&82&0.90&2.1&&67&0.92&12&0.54&0.01&2.7&-0.92&-0.25&1.3&0.58&1.03&1.5\\
7620&93&0.89&1.7&&79&0.92&12&0.55&0.01&2.9&-0.99&-0.10&1.2&0.59&1.08&1.4 
\enddata
\tablecomments{$f_{\rm shell}$ and $R$ are the flux density and radius of the
thin spherical shell. The point source in the east has a flux density $f_1$
and is offset by $\delta{\rm RA}_1$ and $\delta{\rm Dec}_1$ from the shell center,
in RA and Dec respectively. Similarly, $f_2$, $\delta{\rm RA}_2$ and
$\delta{\rm Dec}_2$ are for the point source in the west.}
\tablenotetext{\dag}{The flux ratio between the eastern and western lobes is
defined as $f_{\rm ratio}=\case{f_{\rm shell}/2+f_1}{f_{\rm shell}/2+f_2}$.}
\tablenotetext{\S}{The offset from the shell center to the point source 
is defined as $d_i=\sqrt{(\delta {\rm RA}_i)^2 + (\delta {\rm Dec}_i)^2}$.}
\end{deluxetable}

\end{document}